\documentclass[prl,twocolumn,amsmath,amssymb,superscriptaddress]{revtex4-2}
\usepackage{graphicx}
\usepackage{subfigure}
\usepackage{adjustbox}
\usepackage{physics}
\usepackage{bm}
\usepackage{color}
\usepackage{needspace}
\usepackage{capt-of}
\usepackage{braket}
\usepackage{standalone}
\usepackage{multirow}
\usepackage{tikz}
\usepackage{mathrsfs}
\usepackage{dsfont}
\usepackage{amsfonts}
\usepackage[english]{babel}
\usepackage{dcolumn}
\usepackage{times}

\usepackage[hidelinks,bookmarks=true]{hyperref}

\newcommand{\bea}{\begin{eqnarray}}
\newcommand{\eea}{\end{eqnarray}}

\begin{document}
\title{Exact Parent Hamiltonians for Partially Magnetized Chiral Spin Liquids}

\author{Xin Shen}
\thanks{X. Shen: shencongli.phys@gmail.com}
\affiliation{School of Physics, Peking University, Beijing 100871, China}
\affiliation{Department of Physics, Princeton University, Princeton, NJ 08544, USA}

\author{Guangyue Ji}
\affiliation{International Center for Quantum Materials, Peking University, Beijing 100871, China}
\affiliation{Beijing Key Laboratory of Quantum Devices, Peking University, Beijing, 100871, China}

\author{Jie Wang}
\thanks{J. Wang: jiewang.phy@pku.edu.cn}
\affiliation{International Center for Quantum Materials, Peking University, Beijing 100871, China}
\affiliation{Beijing Key Laboratory of Quantum Devices, Peking University, Beijing, 100871, China}

\begin{abstract}
    Chiral spin liquids (CSLs) are paradigmatic fractionalized phases, yet simple exact parent Hamiltonians for them remain scarce. In this work, we construct exact spin parent Hamiltonians for partially magnetized CSLs realizing various bosonic topological orders. Our construction relies on an exact Hubbard-type parent Hamiltonian for bosonic lattice fractional quantum Hall (FQH) states, valid at arbitrary repulsive interaction strengths. Taking the on-site repulsion to infinity and applying local boson–spin mappings yields exact parent Hamiltonians for the CSLs. Our two-spin construction generalizes the Kalmeyer–Laughlin spin Hamiltonian to a broader class of bosonic FQH model states, including Abelian states and non-Abelian parton states. We derive controlled microscopic realizations of these two-spin Hamiltonians via second-order superexchange in generalized spinful Hofstadter--Hubbard models with tailored hoppings and interactions. With more general local constraints and interactions, boson--spin mappings yield exact higher-spin and multispin parent Hamiltonians for partially magnetized CSLs.
\end{abstract}

\maketitle

Chiral spin liquids (CSLs) are long-range-entangled spin phases with broken time-reversal and mirror symmetries, fractionalized excitations, and chiral edge modes~\cite{KalmeyerLaughlin1987,Kalmeyer1989,WenWilczekZee1989,Wen1990,RMP_YiZhou17}. 
The close connection between CSLs and fractional quantum Hall (FQH) physics~\cite{Laughlin1983,Kalmeyer1989,Panfilov2016} has motivated the search for microscopic realizations of topological order in quantum magnets~\cite{Gang_Review14,He2014,Bauer2014,Gong2014}.

One way to pursue this search is to construct microscopic models with exact spin-liquid ground states. 
Kalmeyer and Laughlin (KL)~\cite{KalmeyerLaughlin1987,Kalmeyer1989} proposed a spin-liquid trial state based on the bosonic Laughlin state at filling \(\nu_b=1/2\) for the nearest-neighbor spin-\(1/2\) triangular-lattice Heisenberg antiferromagnet. Subsequent numerical studies, however, found evidence for $120^\circ$ magnetic order in this model~\cite{BernuLhuillierPierre1992SignatureNeel,CapriottiTrumperSorella1999LongRangeNeel}. This raises the question of which spin Hamiltonians have the KL state as an exact ground state.


{Previous work has constructed parent Hamiltonians for the KL CSL and related spin liquids using multispin exchange~\cite{Schroeter2007,ThomaleKapitSchroeterGreiter2009ParentCSL,NielsenCiracSierra2012,GreiterSchroeterThomale2014ParentNACSL}, as well as for lattice FQH states~\cite{GlasserCiracSierraNielsen2015,Manna2018Quasiholes,Manna2019Quasielectrons,Manna2020Fractal,JiZhang2026LatticeFQH}.
One limitation of the original KL model is the absence of a lattice Landau-level subspace.
The Kapit--Mueller (KM) model~\cite{KapitMueller2010} realizes a lattice lowest Landau level (LLL) in which the bosonic Laughlin state at filling $\nu_b=1/2$ is an exact ground state for repulsive on-site interactions.
In the hard-core limit, the partially magnetized KL CSL can be realized by a Hamiltonian with only two-spin XY exchange.
The same approach extends to $Z_k$ Read--Rezayi-type CSLs based on exact bosonic FQH states at filling $\nu_b=k/2$ stabilized by effective $(k+1)$-body on-site repulsions~\cite{KapitSimon2013}.}

In this work, we construct spin Hamiltonians for more general Abelian and non-Abelian CSL model states in lattice conformal Hilbert spaces, including the lattice-sampled LLL and beyond. We first discuss the KM-based XY model stabilizing the KL spin liquid and its Hubbard realization through superexchange. Taking this model as a starting point, we extend the construction along two routes. In the first, we modify the two-spin exchange couplings to access higher-rank flatband subspaces or bands with higher Chern number. In the second, we retain the KM exchange and add Ising-type interactions corresponding to products of boson densities on distinct sites to obtain deformed Laughlin and dressed Moore--Read spin states at other fillings within the lattice LLL. The key to our construction is an exact parent Hamiltonian for the bosonic FQH ground state on a lattice, valid at arbitrary repulsion strengths; taking the on-site component of interaction to infinity maps our boson model to spin Hamiltonians. Our work deepens the understanding of the intrinsic connections among FQH, Hubbard, and spin-liquid physics and suggests potential routes to realizing topological order in engineered platforms.

\paragraph{Exact partially magnetized KL CSL.}
We begin with the exact spin Hamiltonian for the KL CSL
obtained from hard-core bosons with on-site repulsion at filling
$\nu_b = 1/2$ in the lattice LLL~\cite{KapitSimon2013}. 
We emphasize that the parent Hamiltonian realized through this route is not spin $SU(2)$ symmetric and its exact KL CSL has
nonzero magnetization, hence should be termed as partially magnetized CSL~\cite{Shen2018Magnetized,PanYangChen2025}.
The advantage of the KM-based spin Hamiltonian is that only
two-spin interactions are required, making its exchange structure much simpler.

The KM model is a fine-tuned Hofstadter model. Denote the Bravais lattice as $\Lambda = \{m\bm a_1+n\bm a_2:m,n\in\mathbb Z\}$, with primitive lattice vectors $\bm a_{1,2}$ spanning an area $|\bm a_1\times\bm a_2| = 2\pi\ell^2$. The KM model is:
\begin{gather}
    H_{\rm KM} = \sum_{\bm r,\bm d\in \Lambda} W_{\bm d}\, e^{i\varphi_{\bm r,\bm d}}\, c^\dagger_{\bm  r+\bm d} c_{\bm r}, \label{eq:HKM_def}\\
    W_{\bm d} = J e^{-(1-\phi)|\bm d|^2/4\ell^2},\quad \varphi_{\bm r,\bm d} = (\phi+1)\bm r\times\bm d/2\ell^2. \nonumber
\end{gather}
Here $c_{\bm r}$ are canonical boson operators with unrestricted
onsite occupation, $J>0$, and $\phi\in(0,1)$ is the flux per
primitive cell in units of the flux quantum.
We use the convention
$\bm r\times\bm d = (\bm r\times\bm d)\cdot\hat{\bm z}$.
Compared with the original KM model~\cite{KapitMueller2010},
we include the onsite term and incorporate the $Z_2$ hopping-sign
structure into the Peierls phase, giving positive real hopping
prefactors and an effective flux $\phi+1$ per primitive cell.
In the above model, the flatband has exact zero energy.

For $\phi = p/q$ with coprime $p,q$, the lowest $p$ bands form the
zero-energy sampled-LLL manifold~\cite{KapitMueller2010}. Adding a two-body onsite repulsion to the KM hopping gives
\begin{equation}
    H_{\rm KM}^{(2)}=H_{\rm KM}+V, \quad V = \frac{U}{2} \sum_{\bm r} n_{\bm r} (n_{\bm r}-1), \label{eq:HBH_def}
\end{equation}
where $U > 0$ represents the strength of on-site repulsion and $n_{\bm r} = c_{\bm r}^{\dagger}c_{\bm r}$ is boson density. For $N$ bosons on $N_s$ sites, $\nu_b=N/(\phi N_s)$. At $\nu_b=1/2$,
let $|\Phi_{\rm L}\rangle$ denote the lattice-sampled Laughlin state in the site gauge of Eq.~\eqref{eq:HKM_def}, with amplitudes
$\Phi_{\rm L}(\{z\})\prod_{i=1}^{N}\eta(z_i)$. Here $\{z\}:=\{z_1,\ldots,z_N\}$, $z_i=r_{i,x}+ir_{i,y}$, and $\eta(z_i) := \eta_{\bm r_i}=(-1)^{m_i+n_i+m_in_i}$, with $\bm r_i = m_i\bm a_1+n_i\bm a_2$. On the infinite plane,
\begin{equation}
    \Phi_{\rm L}(\{z\}) \propto \prod_{i<j}(z_i-z_j)^2 e^{-\frac{\phi}{4\ell^2}\sum_{i=1}^{N}|z_i|^2}, \label{eq:PhiL_continuum}
\end{equation}
whose torus geometry extension is straightforward~\cite{HaldaneRezayi1985PeriodicLaughlin,Haldane2018ModifiedWeierstrassSigma} (see Sec.~\ref{app:KM_gauge} for the gauge and torus conventions).

Following Ref.~(\onlinecite{KapitMueller2010}), we emphasize that the bosonic Laughlin $\nu_b = 1/2$ state is the exact zero-energy ground state of $H_{\rm KM}^{(2)}$ for any strength of on-site repulsive interactions. This is a consequence of the sampled-LLL structure ($H_{\rm KM}|\Phi_{\rm L}\rangle=0$) and the quadratic coincidence zeros ($V|\Phi_{\rm L}\rangle=0$), together with the positive semidefiniteness of both terms. Thus, $H_{\rm KM}^{(2)}$ is frustration-free in the sense that its kinetic and interaction energies are minimized simultaneously. Based on this insight, taking the hard-core boson limit $U \to \infty$ yields an effective spin-$1/2$ local Hilbert space and transforms $H_{\rm KM}^{(2)}$ into a spin model whose ground state is a KL-type CSL~\cite{KapitSimon2013}. Concretely, denoting the on-site basis states by $|0\rangle_{\bm r}$ and $|1\rangle_{\bm r}$ and the hard-core boson creation operator as $b_{\bm r}^{\dagger} = |1\rangle_{\bm r}\langle0|$, the Matsubara--Matsuda map~\cite{MatsubaraMatsuda1956}
turns bosonic operators into spin-$1/2$ operators,
\begin{equation}
    b^\dagger_{\bm r}\leftrightarrow S^+_{\bm r},\quad b_{\bm r}\leftrightarrow S^-_{\bm r},\quad n_{\bm r}\leftrightarrow S^z_{\bm r}+\frac12, \label{eq:MM_map}
\end{equation}
and the hard-core restriction of Eq.~\eqref{eq:HKM_def} becomes
\begin{equation}
    H_{\rm XY} = \sum_{\bm r,\bm d}W_{\bm d}\,e^{i\varphi_{\bm r,\bm d}}\, S^+_{\bm r+\bm d}S^-_{\bm r}. \label{eq:H_XY_intro}
\end{equation}
Consequently, the bosonic Laughlin state maps to the KL spin state,
\begin{equation}
    |\Psi_{\rm KL}\rangle = \sum_{\{z\}} \Phi_{\rm L}(\{z\}) \prod_{i=1}^{N} \bigl[\eta(z_i)S^+_{z_i}\bigr] |\Downarrow\rangle, \label{eq:Psi_KL_spin}
\end{equation}
where $S^+_{z_i}:=S^+_{\bm r_i}$, $|\Downarrow\rangle$ is the
all-spin-down state, and each $N$-site subset is counted once.
The exact mapping gives a zero-energy ground state of $H_{\rm XY}$
in the sector $M=S^z_{\rm tot}=N-N_s/2$. With saturation magnetization $M_s=N_s/2$, its normalized magnetization is
$m_0=M/M_s=2\phi\nu_b-1=\phi-1$ at $\nu_b=1/2$. 
We notice that since the spin Hamiltonian conserves $S^z_{\rm tot} = \sum_{\bm r}S^z_{\bm r}$, the ground-state magnetization can be tuned by applying a uniform Zeeman term; this applies to all spin models discussed in this work as they all conserve total $S^z$. For the two-spin models, reversing the net Zeeman field at every site maps a CSL to its time-reversed partner with opposite magnetization and chirality (Sec.~\ref{app:TR_symmetry}).

\begin{figure}[!t]
    \centering
    \setlength{\abovecaptionskip}{3.5pt}
    \setlength{\belowcaptionskip}{0pt}
    \includegraphics[width=\columnwidth]{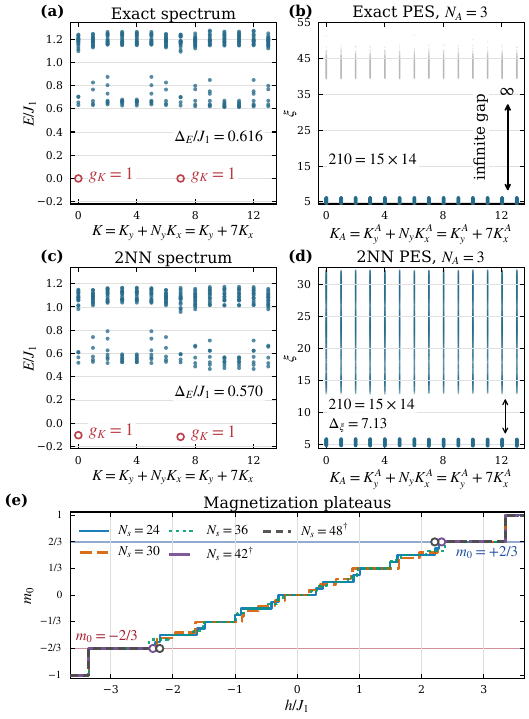}
    \caption{\textbf{The KL spin Hamiltonian and its second-neighbor truncation on triangular tori.} (a)--(d): $6\times7$, $N_s=42$, $N=7$, $\phi=1/3$. (a,c) Exact and 2NN spectra in units of the nearest-neighbor
hopping scale $J_1$, showing the lowest 20 levels per momentum sector and an isolated two-state manifold with one state at $K=0$ and one at $K=7$. In (a,c), $\Delta_E$ is the gap above the two-state manifold. (b,d) PES: 15 levels per momentum sector below the gap, matching Laughlin quasihole counting~\cite{BernevigHaldane2008Jack}. Gray points in (b) are numerical zeros and correspond to infinite entanglement energies in exact arithmetic. The upper branch in (d) is finite. (e) Ground-state magnetization $m_0=2\langle S^z_{\rm tot}\rangle/N_s$ of the finite-torus KM spin Hamiltonian $H_{\rm XY}-(\bar\epsilon+h)S^z_{\rm tot}$ at $\phi=1/3$, where $\bar\epsilon$ is the mean coefficient of the onsite terms \(S_{\boldsymbol r}^{+}S_{\boldsymbol r}^{-}\) in \(H_{\rm XY}\). At each field, the ground state is selected by comparing the lowest energies of the magnetization sectors. The red and blue guides mark the KL plateaus at $m_0=-2/3$ and $+2/3$, respectively. Complete curves are shown for $N_s=24,30,36$. For $N_s=42,48$ ($\dagger$), only the KL plateaus and fully polarized branches are shown.} \label{fig:kl-numerics}
\end{figure}

To examine the excitation gap and topological order of the model
in Eqn.~(\ref{eq:H_XY_intro}), many-body numerics is indispensable. On triangular tori, the finite-size spectra and particle entanglement spectra (PES) are consistent with Laughlin order [Fig.~\ref{fig:kl-numerics}(a,b)]. For the long-range XY model at $\phi=1/3$, the magnetization curves
exhibit plateaus at $m_0=\pm2/3$, allowing field selection of
opposite-chirality sectors [Fig.~\ref{fig:kl-numerics}(e)].

On the studied torus, truncating the Gaussian-decaying KM exchanges to second neighbors (2NN) preserves the separated two-state low-energy manifold and Laughlin PES counting~\cite{LiHaldane2008,SterdyniakRegnaultBernevig2011,
SterdyniakRegnaultMoller2012} below a clear entanglement gap
[Fig.~\ref{fig:kl-numerics}(c,d)], supporting finite-range implementations. These finite-range XY models are candidates for implementation in programmable superconducting circuit-QED arrays with photon blockade and synthetic gauge fields~\cite{WangPhotonFQH2024,Rosen2024SyntheticVectorPotential}.

The construction of the KL spin Hamiltonian extends from Bravais to non-Bravais lattices~\cite{DongMueller2020ExactFlatBands,LiWu2026LatticeGLL} (Sec.~\ref{app:nonbravais_spin}). Bosonic $Z_k$ Read--Rezayi states at filling $\nu_b=k/2$~\cite{ReadRezayi1999} also provide starting points for spin models~\cite{KapitSimon2013}. In the hard-core limit of the on-site $(k+1)$-body repulsion, the allowed occupation states $n=0,\ldots,k$ correspond one-to-one to the $S^z$ eigenstates of a local spin $S=k/2$, as discussed in Sec.~\ref{app:spin_truncated_boson}.

\paragraph{Hubbard realization and superexchange.}
To connect the XY Hamiltonian \eqref{eq:H_XY_intro} to Bose-- and Fermi--Hubbard realizations, we derive it from a spinful Bose--Hubbard model through second-order superexchange. The corresponding Fermi--Hubbard realization is given in Sec.~\ref{app:fermi_hubbard}.
For a Bravais lattice, let $\mathcal D_+\subset\Lambda\setminus\{\bm0\}$ contain one representative of each pair $\{\bm d,-\bm d\}$. We consider the following Hamiltonian:
\begin{equation}
\begin{aligned}
    H &= H_t+H_U+H_Z,\qquad H_Z = -h \sum_{\bm r} S^z_{\bm r},\\
    H_t
    &= -\sum_{\bm r,\sigma}\sum_{\bm d\in\mathcal D_+} t_{\bm d} ~ e^{i s_\sigma\theta_{\bm r,\bm d}} ~ a^\dagger_{\bm r+\bm d,\sigma} a_{\bm r,\sigma} + \mathrm{H.c.} ,\\
    H_U
    &= \frac{1}{2} \sum_{\bm r,\sigma} ~ U_\sigma ~ n_{\bm r,\sigma} (n_{\bm r,\sigma} - 1) + U_{\uparrow\downarrow} \sum_{\bm r} n_{\bm r,\uparrow} n_{\bm r,\downarrow},
\end{aligned} \label{eq:H_Hubbard_SU2}
\end{equation}
where $a_{\bm r,\sigma}$ are canonical boson operators, $n_{\bm r,\sigma} = a^\dagger_{\bm r,\sigma}a_{\bm r,\sigma}$ is the local density operator, and $S_{\bm r}^a = a^\dag_{\bm r,s} \sigma^a_{s,s'} a_{\bm r,s'}/2$ is the local spin operator. On the bond from $\bm r$ to $\bm r+\bm d$, $t_{\bm d}$ is the real hopping amplitude and $s_\sigma\theta_{\bm r,\bm d}$ is the Peierls phase, with $s_\uparrow=+1$ and $s_\downarrow=-1$. On the reverse bond, $\theta_{\bm r+\bm d,-\bm d}=-\theta_{\bm r,\bm d}$. The onsite repulsions are $U_\sigma$ for two bosons of the same spin $\sigma$ and $U_{\uparrow\downarrow}$ for two bosons of opposite spin. The model conserves total $S^z$.

In the unit-filled Mott regime, the singly occupied states form a spin-$1/2$ doublet, with $S_{\bm r}^+ = a^\dagger_{\bm r,\uparrow}a_{\bm r,\downarrow}$. The exchange part of the second-order effective Hamiltonian is~\cite{DuanDemlerLukin2003,SchriefferWolff1966},
\begin{equation}
  H_{\rm ex}=\frac{1}{2}\sum_{\bm r,\bm d\neq\bm0}
  \left[J_{\bm d}^{z}S_{\bm r+\bm d}^{z}S_{\bm r}^{z}
  +J_{\bm d}^{xy}e^{i2\theta_{\bm r,\bm d}}
  S_{\bm r+\bm d}^{+}S_{\bm r}^{-}\right],
  \label{eq:H_exchange_general}
\end{equation}
where $J_{\bm d}^{z}=4|t_{\bm d}|^2(U_{\uparrow\downarrow}^{-1}-U_\uparrow^{-1}-U_\downarrow^{-1})$ and $J_{\bm d}^{xy}=-4|t_{\bm d}|^2/U_{\uparrow\downarrow}$. The sum in Eq.~\eqref{eq:H_exchange_general} runs over both bond
orientations. To reproduce the KM exchange, we choose $U_\uparrow=U_\downarrow=2U_{\uparrow\downarrow}\equiv U_0$. For $\bm d\in\mathcal D_+$, we set $t_{\bm d}=t_0e^{-(1-\phi)|\bm d|^2/(8\ell^2)}$ and $\theta_{\bm r,\bm d}=(\phi+1)\bm r\times\bm d/(4\ell^2)+\pi/2$. Then $J_{\bm d}^{z}=0$ and $e^{i2\theta_{\bm r,\bm d}}=-e^{i\varphi_{\bm r,\bm d}}$, so
\begin{equation}
    H_{\rm ex} = \sum_{\bm r,\bm d\neq\bm0}
    W_{\bm d}\,e^{i\varphi_{\bm r,\bm d}}\,S^+_{\bm r+\bm d} S^-_{\bm r}, \label{eq:Hex_equals_HXY}
\end{equation}
which is the $\bm d\neq\bm0$ part of $H_{\rm XY}$, with \(J=4t_0^2/U_0\) to second order in hopping. The uniform onsite contribution is supplied by $H_Z$, up to an additive constant. Higher-order corrections become negligible compared with $J$ as
$t_0/U_0\to0$ (Sec.~\ref{app:bose_hubbard}).

For bosons, the $\pi/2$ offset compensates the negative
superexchange sign. Laser-assisted tunneling in spin-dependent Hofstadter optical lattices produces opposite Peierls phases for two internal states~\cite{Aidelsburger2013}; related experiments demonstrate control of hopping phases~\cite{Miyake2013}. Several phase-locked, displacement-resolved tunneling channels could approximate a finite-range truncation of the Gaussian hopping profile.

\paragraph{Exact spin Hamiltonians beyond the KL CSL.} In the preceding discussion, we derived the parent spin Hamiltonians for the Laughlin $\nu_b = 1/2$ CSL. We started from the lattice frustration-free parent Hamiltonian for the bosonic Laughlin $1/2$ topological order. The on-site interaction of infinite strength yields a hard-core boson local Hilbert space which enables a mapping to a spin-$1/2$ problem. The hopping of bosons then translates into spin exchanges.

Applying the same local boson-to-spin mapping to known frustration-free bosonic parent Hamiltonians in lattice conformal Hilbert spaces, we extend the KL construction to (1) color-entangled Halperin $(221)$ topological order in a higher-Chern flatband; (2) Abelian and non-Abelian parton states in a higher-rank manifold of degenerate flat bands; and (3) deformed Laughlin and dressed Moore--Read states in the lattice LLL. Cases (1) and (2) share the same two-body on-site interaction but differ in their hopping. In (3), the hopping retains the KM form, while the interactions extend beyond two-body on-site repulsion. We discuss their spin parent Hamiltonians and corresponding magnetization sectors. These constructions may also be combined to explore richer CSL topological orders, such as non-Abelian spin-singlet states stabilized by few-body interactions in a higher Chern band~\cite{ArdonneReadRezayiSchoutens2001NASS,BehrmannLiuBergholtz2016}. For selected extensions, Appendix~\ref{app:numerics} presents finite-size gaps above the zero-energy ground states and momentum-resolved PES counting consistent with exclusion rules or independent model-state calculations.

\emph{Route 1: Parton and Halperin spin-liquid states.}
In this part, we consider topological orders still stabilized by two-body on-site interactions but with modified hopping terms. The modification includes either a higher Chern number or a higher rank conformal Hilbert space. The lattice conformal Hilbert spaces for the parton and Halperin constructions are built from a set of degenerate flat bands and a single higher-Chern band, respectively. The corresponding spin-$1/2$ model retains the two-spin exchange form, with modified couplings.

The general form of this type of spin Hamiltonian is:
\begin{equation}
    H_{\rm XY}^{(\mathrm P)} = \sum_{\bm r,\bm d}W_{\bm r,\bm d}^{(\mathrm P)} e^{i\varphi_{\bm r,\bm d}^{(\mathrm P)}} S_{\bm r+\bm d}^{+}S_{\bm r}^{-}. \label{eq:Hspin_general_nLL}
\end{equation}

In the above, $(\mathrm P)$ labels the projector construction. $h$ is its positive-semidefinite single-particle Hamiltonian, with $W_{\bm r,\bm d}^{(\mathrm P)} e^{i\varphi_{\bm r,\bm d}^{(\mathrm P)}} = \langle\bm r+\bm d|h|\bm r\rangle$ and $W_{\bm r,\bm d}^{(\mathrm P)} \ge 0$. The sum includes all directed pairs of physical sites as well as the onsite terms $\bm d = \bm0$; on non-Bravais lattices, $\bm r$ denotes the lattice-site position. Below, we describe the topological order and the magnetization sector at the corresponding boson filling for different Hamiltonians $h$ constructed from projectors. We consider two projector constructions~\cite{CruiseSeidel2026,LiWu2026LatticeGLL}. One selects a higher-rank subspace spanned by degenerate lattice-sampled Landau levels, and the other selects a single ideal higher-Chern flatband. The same projector construction extends to other single-particle subspaces supporting hard-core bosonic model states.

For the higher-rank case, we extend the explicit magnetic Bloch formulation of Ref.~\cite{CruiseSeidel2026} from $\phi=1/q$ to
general rational flux $\phi=p/q$, including $p>1$. Let $P_{\bm k}^{\parallel}$ project onto the first
$n_{\rm LL}$ sampled LLs. We choose
\(h_{\boldsymbol k}^{(n_{\rm LL})}=E_{\boldsymbol k}(I-P_{\boldsymbol k}^{\parallel})\), with \(E_{\boldsymbol k}>0\). The sampled LL subspace has zero energy, while the remaining bands share the dispersion \(E_{\boldsymbol k}\). For $pn_{\rm LL}<q$ and independent sampled states throughout the magnetic Brillouin zone, the sampled LL subspace has dimension \(pn_{\rm LL}\) at each magnetic Bloch momentum (its rank) and total Chern number \(n_{\rm LL}\). The normalization of the sampled LL orbitals and the positive dispersion $E_{\bm k}$ are specified in Sec.~\ref{app:projector_mll}. The corresponding spin Hamiltonian is denoted by $H_{\rm XY}^{(n_{\rm LL})}$. The associated model states have the parton form,
\begin{equation}
    \Psi_{mn}(\{z_i\}) = \chi_m(\{z_i\})\chi_n(\{z_i\}).
    \label{eq:parton_chim_chin}
\end{equation}
Here $\chi_\ell$ is a fermionic integer-QH Slater determinant with $\ell$ filled Landau levels, evaluated at the lattice coordinates. The product $\chi_m\chi_n$ gives the amplitude for the spin configuration
with its $N$ up spins at $\{z_i\}$. With the two effective fields chosen to sum to the physical field, the unprojected product lies in the first $n_{\rm LL}=m+n-1$ LLs (Sec.~\ref{app:projector_mll}). It is bosonic and vanishes at least quadratically at coincidence, giving an exact zero mode at $\nu_b=mn/(m+n)$~\cite{CruiseSeidel2026}.

Examples are the Abelian Jain-type $\chi_1\chi_2$ state, the Ising-type $\chi_2^2$ state, and the Fibonacci-type candidate $\chi_3^2$, with
$(\nu_b,n_{\rm LL})=(2/3,2),(1,3),(3/2,5)$, respectively~\cite{Jain1989,Wen1991,Ahari2023Fibonacci,Bandyopadhyay2018EPP,Bandyopadhyay2020Jain,CruiseSeidel2023DNA,CruiseSeidel2026}. Their magnetization follows from $m_0=2\phi\nu_b-1$. We present the finite-size energy spectra and PES for $p=1$ and $p>1$ (Appendix~\ref{app:numerics}, Figs.~\ref{fig:parton_supp} and~\ref{fig:parton_rational}, respectively). We also derive a counting rule for the root states of these parton models using the entangled Pauli principle (EPP). For the $\chi_3^2$ model, we conjecture that every root state allowed by the local constraints can be completed to a many-body zero mode. On the finite torus studied here, the periodic version of the rule reproduces the quasihole zero-mode count in each momentum sector (Fig.~\ref{fig:h33_n4_hardcore}).

We also construct spin-$1/2$ parent Hamiltonians from a single ideal higher-Chern band. In the non-Bravais family of Ref.~\cite{LiWu2026LatticeGLL} with $n_{\rm sub}\ge3$ physical sites per primitive cell, let $Q_{\bm k}^{\rm LL}$ project onto the first $n_{\rm sub}-1$ modulated and sampled LLs. Its one-dimensional orthogonal complement is an ideal band with $C=-(n_{\rm sub}-1)$~\cite{LiWu2026LatticeGLL}. We choose $h_{\bm k}^{(C)}=E_{\bm k}Q_{\bm k}^{\rm LL}$, with $E_{\bm k}>0$, placing the higher-Chern band at zero energy and the sampled-LL subspace at positive energy.

Ideal higher-Chern bands admit a color-entangled description in terms of multicomponent Landau levels~\cite{WuRegnaultBernevig2013,WangKlevtsovLiu2023Ideal,DongLedwith2023Ideal}. Bosonic Halperin-type states occur in model higher-Chern bands with two-body on-site repulsion at filling $\nu_b=N/N_{\rm cell}=1/(|C|+1)$~\cite{BehrmannLiuBergholtz2016,WangKlevtsovLiu2023Ideal}, where $N_{\rm cell}$ is the number of primitive cells. For the spin model constructed from the $C=-2$ band, the finite-size energy spectrum and PES level counts at $\nu_b=1/3$ are consistent with Halperin $(221)$ topological order (Appendix~\ref{app:numerics}, Fig.~\ref{fig:highchern54-spectrum-pes}).

For both constructions, Sec.~\ref{app:projector_mll} gives the conditions for a single-particle gap and exponentially decaying hopping~\cite{ChenMazaheriSeidelTang2014,LeeArovasThomale2016Flatness}. Their Hubbard realizations are discussed in Secs.~\ref{app:fermi_hubbard} and~\ref{app:bose_hubbard}.

\emph{Route 2: Generalized Read-Rezayi spin-liquid states.} We construct spin-$1/2$ parent Hamiltonians for deformed Laughlin and dressed Moore--Read states, following the bosonic lattice construction of Ref.~\cite{JiWang2026}. We retain the KM exchange $H_{\rm XY}$ and add Ising-type interactions corresponding to products of boson densities on distinct sites. With a $(k+1)$-site interaction for $k\ge2$, the Hamiltonian is
\begin{equation}
    \begin{aligned}
    H_{\rm def}^{(k,n_2)}
    &=H_{\rm XY}+H_{2,\delta}^{(n_2)}+H_{k+1,\delta'},\\
    H_{2,\delta}^{(n_2)}
    &=\sum_{\bm r}\sum_{\lambda=1}^{n_2}
    \mathcal V_{\bm\delta_\lambda}^{(2)}
    \left(S_{\bm r}^{z}+\frac12\right)
    \left(S_{\bm r+\bm\delta_\lambda}^{z}+\frac12\right),\\
    H_{k+1,\delta'}
    &=\mathcal V_{\bm\delta'}^{(k+1)}
    \sum_{\bm r}\prod_{\ell=0}^{k}
    \left(S_{\bm r+\ell\bm\delta'}^{z}+\frac12\right).
    \end{aligned} \label{eq:spin_half_deformed_parent}
\end{equation}
Here $n_2$ counts distinct nonzero pair displacements
$\bm\delta_\lambda$ up to inversion, and $\bm\delta'$ is the cluster
spacing. All interaction coefficients are positive, and each cluster
contains distinct sites. The spin-model couplings $\mathcal V$ are
distinct from the microscopic Hubbard interactions $V_{\bm d}^{s,o}$. A nonzero hard-core bosonic state whose single-particle support lies in the zero-energy subspace and which satisfies these density constraints gives an exact spin zero mode.

With only the two-site density interactions,
$H_{\rm XY}+H_{2,\delta}^{(n_2)}$ admits the deformed Laughlin sequence at $\nu_b=1/[2(n_2+1)]$. A $k=1$ cluster term is itself a pair interaction and is included among these $n_2$ displacements.
The $\nu_b=1/4$ member has been studied numerically through its
spectrum and PES in the KM flat band~\cite{JiWang2026}. After we absorb the uniform linear $S^z$ terms into the net
Zeeman field, the XY exchange and the Ising interaction are both invariant under time reversal. The resulting exchange Hamiltonian admits an extended-Hubbard realization from superexchange 
(Secs.~\ref{app:fermi_hubbard} and~\ref{app:bose_hubbard}).

For $k=2$ and even $N$, we construct a generalized Jastrow-dressed Moore--Read
state on the lattice. The polynomial part of this state can also be constructed by multiplying the clustering polynomial derived from Macdonald polynomials by a Jastrow factor~\cite{JiWang2026}. Without the additional pair constraints,
its polynomial part is
\begin{equation}
    \mathcal P_{\rm dMR}(\{z_i\}) = \operatorname{Pf} \! \left[\frac{z_{ij}}{z_{ij}^{2}-(\delta')^{2}}\right] \prod_{i<j} z_{ij} \bigl[z_{ij}^{2}-(\delta')^{2}\bigr]. \label{eq:dressed_MR_main}
\end{equation}
Here $z_{ij}=z_i-z_j$, and $\delta'$ is the complex coordinate of
the cluster spacing $\bm\delta'$. The Pfaffian denominators cancel
against the product. The resulting polynomial has at least quadratic
coincidence zeros and vanishes whenever three particles occupy
$z,z+\delta',z+2\delta'$. For nonzero $n_2$, multiplication by
$\prod_{i<j}\prod_{\lambda=1}^{n_2}(z_{ij}^{2}-\delta_\lambda^{2})$,
with $\delta_\lambda$ the complex coordinate of $\bm\delta_\lambda$,
imposes the additional pair constraints. With the Gaussian and
site-gauge factors used in Eq.~\eqref{eq:Psi_KL_spin}, these polynomials
give the spin amplitudes of exact zero-energy ground states of $H_{\rm def}^{(2,n_2)}$.
Sec.~\ref{app:positive_delta_interactions} gives the full amplitudes and proof. For $n_2=0$ at $\nu_b=1/3$, the finite-torus spectrum contains nine zero modes, and their PES level counts agree with the prediction for the dressed Moore--Read state (Appendix~\ref{app:numerics}, Fig.~\ref{fig:rr_supp}(e,f)).

\paragraph{Discussion and Outlook.}
We have constructed exact spin parent Hamiltonians and derived
controlled second-order Hubbard realizations for the two-spin models.
Numerical evidence for the KL phase with exchanges truncated at second
neighbors motivates finite-range implementations. We anticipate spin-dependent optical lattices, superconducting circuits, and platforms with programmable higher-body interactions offer complementary implementation routes for realizing our models~\cite{Aidelsburger2013,Miyake2013,WangPhotonFQH2024,
Katz2023,Weimer2010,Samajdar2026}.
In particular, superconducting processors realize two-dimensional
XY models with spatially and temporally programmable couplings
and local Zeeman fields
\cite{Liu2026Prethermalization,Andersen2025Thermalization}. Further directions include explicit wavefunctions for higher-$k$
displaced-cluster constraints and multicomponent spin-liquid
constructions
\cite{MooreRead1991,ReadRezayi1999,
GreiterSchroeterThomale2014ParentNACSL,YaoKivelson2007}. The exact mapping between the boson and spin Hamiltonians also motivates studying spin-liquid counterparts of collective excitations~\cite{GirvinMacDonaldPlatzman1986,LiouHaldaneYangRezayi2019,XavierBacciconiChandaSonDalmonte2025}. It would also be interesting to explore nearby phases, including possible anyon superfluidity~\cite{Laughlin1988,WenWilczekZee1989}, by varying the magnetization or doping the underlying Hubbard models.

\paragraph{Acknowledgments.}
We thank Andrei Bernevig, Gang Chen, Duncan Haldane, Biao Lian, Rui-Heng Liu, Ying-Hai Wu, and Wei Zhu for valuable discussions. We acknowledge the use of Codex (OpenAI; GPT-5.6 Sol, GPT-6 Astra, and GPT-6.1 Sol) for discussions and assistance in manuscript organization.

This work was supported by the National Natural Science Foundation of China (Grant No. 12674196), the Quantum Science and Technology-National Science and Technology Major Project (Grant No. 2025ZD0300500), and the Fundamental Research Funds for the Central Universities, Peking University. X.S. acknowledges the hospitality of Peking University and Princeton University, where part of this work was completed.

\bibliography{CSL_B3_WORKING}

\appendix
\onecolumngrid

\setcounter{secnumdepth}{2}
\makeatletter
\renewcommand\subsection{\@startsection{subsection}{2}{\z@}
 {-2.2ex plus -.3ex minus -.2ex}{1.1ex plus .2ex}
 {\normalfont\normalsize\bfseries}}

\renewcommand\thesubsection{\thesection.\arabic{subsection}}
\renewcommand\p@subsection{}
\makeatother

\newpage
\section*{------ APPENDIX ------}

\medskip
\begingroup
\setlength{\parindent}{0pt}

\newcommand{\SMentry}[3][]{\par\noindent \if\relax\detokenize{#1}\relax\else\fi\hyperref[#2]{\makebox[2.8em][l]{\ref*{#2}}#3}\nobreak\leaders\hbox{\kern.25em.\kern.25em}\hfill\nobreak\hyperref[#2]{\pageref*{#2}}\par\vspace{3pt}}

\newcommand{\SMmajor}[3][]{\par\vspace{5pt}{\bfseries\SMentry[#1]{#2}{#3}}}

\SMmajor{app:lattice_foundations}{Kalmeyer--Laughlin chiral spin liquid and related lattice models}
\SMentry{app:KL_review}{Kalmeyer--Laughlin construction}
\SMentry{app:nonbravais_spin}{Exact spin Hamiltonians on non-Bravais lattices}
\SMentry{app:KM_gauge}{Gauge conventions and magnetic translations}
\SMmajor[O01]{app:hubbard_symmetry}{Hubbard realizations and time-reversal symmetry}
\SMentry{app:fermi_hubbard}{Fermi--Hubbard realization}
\SMentry{app:bose_hubbard}{Bose--Hubbard realization}
\SMentry[O02]{app:TR_symmetry}{Time-reversal symmetry and Zeeman fields}
\SMmajor{app:generalizations}{Generalized spin Hamiltonians and exact states}
\SMentry[O03]{app:projector_mll}{Multi-LL and higher-Chern constructions}
\SMentry{app:positive_delta_interactions}{Density interactions between distinct sites and exact states}
\SMentry{app:spin_truncated_boson}{Higher-spin Hamiltonians and Read--Rezayi states}
\SMmajor{app:numerics}{Finite-size energy spectra and particle entanglement spectra}
\SMentry{app:numerical_setup}{Numerical setup and observables}
\SMentry{app:pes_counting}{PES and zero-mode counting}
\SMentry{app:numerical_results}{Finite-size spectra and counting comparisons}
\endgroup

Appendix~\ref{app:lattice_foundations} reviews the original proposal by Kalmeyer and Laughlin, gives exact spin Hamiltonians on non-Bravais lattices, and specifies the gauge and boundary conditions used on a torus. Appendix~\ref{app:hubbard_symmetry} derives the spin couplings in Bose-- and Fermi--Hubbard models and explains how a Zeeman field splits the energies of time-reversed CSL states. Appendix~\ref{app:generalizations} constructs spin Hamiltonians for parton states in multiple sampled Landau levels and Halperin-type states in higher-Chern bands. It also covers deformed Laughlin and dressed Moore--Read states, as well as higher-spin Read--Rezayi models. Appendix~\ref{app:numerics} presents finite-size energy spectra and particle entanglement spectra. It also derives an EPP counting rule for parton root states and tests its prediction for the number of quasihole zero modes in the $\chi_3^2$ model.\\

\noindent\textbf{Models and extensions}\par\smallskip
\begingroup
\normalsize
\setlength{\tabcolsep}{4pt}
\renewcommand{\arraystretch}{1.15}
\newcommand{\SMcell}[1]{\raggedright\leavevmode #1}
\noindent\begin{tabular}{@{}p{.18\linewidth}p{.23\linewidth}p{.245\linewidth}p{.295\linewidth}@{}}
\hline
\SMcell{\textbf{Model / location}} & \SMcell{\textbf{What is changed}} & \SMcell{\textbf{Spin Hamiltonian}} & \SMcell{\textbf{Model states}}\tabularnewline
\hline
\SMcell{KM lattice LLL\newline Main text; \ref{app:KM_gauge}}
& \SMcell{Single lattice lowest Landau level (LLL); Gaussian exchange.}
& \SMcell{$H_{\rm XY}$, Eq.~\eqref{eq:H_XY_intro}\newline $S=1/2$}
& \SMcell{Laughlin/KL\newline $\nu_b=1/2$}\tabularnewline[1ex]
\SMcell{Non-Bravais LLL\newline \ref{app:nonbravais_spin}, \ref{app:KM_gauge}}
& \SMcell{Several sites per primitive cell; Gaussian hopping.}
& \SMcell{$H_{\rm XY}^{G}$, Eq.~\eqref{eq:NB_spin_parent}\newline $S=1/2$}
& \SMcell{Laughlin/KL \newline $\nu_b=1/2$}\tabularnewline[1ex]
\SMcell{Multi-LL parton\newline  \ref{app:projector_mll}, \ref{app:numerics}}
& \SMcell{Retain $n_{\rm LL}=m+n-1$ LLs. At $\phi=p/q$: $pn_{\rm LL}$ states per
  momentum.
  }
& \SMcell{$H_{\rm XY}^{(n_{\rm LL})}$, Eq.~\eqref{eq:Hspin_general_nLL}\newline $S=1/2$}
& \SMcell{Parton $\chi_m\chi_n$\newline $\nu_b=mn/(m+n)$}\tabularnewline[1ex]
\SMcell{Higher-Chern\newline \ref{app:projector_mll}}
& \SMcell{Use the ideal higher-Chern complement of sampled LLs.}
& \SMcell{$H_{\rm XY}^{(C)}$, Eq.~\eqref{eq:HC_sublattice_spin_parent}\newline $S=1/2$}
& \SMcell{Halperin-type\newline $\nu_b=1/(|C|+1)$}\tabularnewline[1ex]
\SMcell{Deformed Laughlin\newline \ref{app:fermi_hubbard}, \ref{app:positive_delta_interactions}}
& \SMcell{Add positive pair repulsions at $n_2$ distinct off-site displacements.}
& \SMcell{$H_{\rm XY}+H_{2,\delta}^{(n_2)}$\newline Eq.~\eqref{eq:spin_half_deformed_parent}; $S=1/2$}
& \SMcell{Deformed Laughlin\newline $\nu_b=1/[2(n_2+1)]$}\tabularnewline[1ex]
\SMcell{Displaced site\newline clusters\newline \ref{app:positive_delta_interactions}, \ref{app:numerics}}
& \SMcell{Add a $(k+1)$-site density interaction, $k\ge2$.}
& \SMcell{$H_{\rm def}^{(k,n_2)}$, Eq.~\eqref{eq:spin_half_deformed_parent}\newline $S=1/2$}
& \SMcell{Dressed Moore--Read ($k=2$)\newline $\nu_b=1/(2n_2+3)$}\tabularnewline[1ex]
\SMcell{Read--Rezayi\newline \ref{app:spin_truncated_boson}, \ref{app:numerics}}
& \SMcell{Single LL; strict onsite occupancy $0,\ldots,k$.}
& \SMcell{$H_{\rm RR}^{(S)}$, Eq.~\eqref{eq:canonical_spinS_parent}\newline $S=k/2$}
& \SMcell{$Z_k$ Read--Rezayi\newline $\nu_b=k/2$}\tabularnewline
\hline
\end{tabular}
\par
\endgroup
\smallskip
\noindent
\phantomsection\label{reorg:OR12}\label{reorg:XR17}\label{sync:O04}

\clearpage
\section{Kalmeyer--Laughlin chiral spin liquid and related lattice models}\label{app:lattice_foundations}
We revisit the original KL proposal for the triangular-lattice antiferromagnet in Sec.~\ref{app:KL_review}. In Sec.~\ref{app:nonbravais_spin}, we extend the spin-$1/2$ construction of the main text to non-Bravais lattices and obtain exact KL ground states. Section~\ref{app:KM_gauge} explains the gauge choice for the KM hopping and identifies the magnetic translations compatible with the boundary conditions on a torus.

\subsection{Kalmeyer--Laughlin construction} \label{app:KL_review}
We review the spin--boson mapping and fictitious-flux interpretation underlying the original KL proposal and explain how the site-gauge transformation enters the trial wavefunction. The nearest-neighbor hopping Hamiltonian yields a dispersive, topologically trivial band, motivating the exact lattice LLL used in the main text.

\paragraph{Spin--boson mapping.}
Kalmeyer and Laughlin (KL) considered the nearest-neighbor spin-$1/2$
triangular-lattice antiferromagnet with exchange $J>0$~\cite{KalmeyerLaughlin1987,Kalmeyer1989}. Write $\bm r=m\bm a_1+n\bm a_2$, with $m,n\in\mathbb Z$, and $\bm a_1=a_0\hat{\bm x}$, $\bm a_2=a_0(\hat{\bm x}+\sqrt3\hat{\bm y})/2$ are two primitive lattice vectors. The area of the unit cell is $A_{\rm uc}=\sqrt3a_0^2/2$.
The three positive nearest-neighbor vectors are
$\bm d_1=\bm a_1$, $\bm d_2=\bm a_2$, and
$\bm d_3=\bm a_2-\bm a_1$.
The map in Eq.~\eqref{eq:MM_map} gives hard-core bosons with
nearest-neighbor hopping $t_0=J/2>0$ and density repulsion $J$.
At zero magnetization, $N=N_s/2$.

\paragraph{Fictitious flux and KL trial state.}
The site transformation $b_{\bm r}\mapsto\eta_{\bm r}b_{\bm r}$,
with $\eta_{\bm r}=(-1)^{m+n+mn}$, expresses the hopping in Peierls form (Sec.~\ref{app:KM_gauge}). For $A_{\rm uc}=2\pi\ell^2$, the complex amplitude associated to tight-binding hopping from site $\bm r$ to site $\bm r+\bm d$, after the transformation, becomes,
\begin{equation}
    t_{\bm r,\bm d}^{\rm KL,\eta}
    =t_0\eta_{\bm r+\bm d}\eta_{\bm r}
    =-t_0e^{i\bm r\times\bm d/(2\ell^2)},
    \qquad \bm d=\pm\bm d_\mu.
    \label{eq:KL_Hofstadter_hopping}
\end{equation}
This indicates that relative to the negative bare hopping, the Peierls phase is $\pi$ per elementary triangle. The full normalized hopping product around a triangle remains
$(-1)^3e^{i\pi}=1$.
With $N_\phi=N_s$, half filling gives
$\nu_b=N/N_\phi=1/2$. In the original spin basis, the corresponding
planar Laughlin trial amplitude has magnetic length
$\ell=\sqrt{A_{\rm uc}/(2\pi)}$ and reads
\begin{equation}
 \Phi_{\rm KL}^{\rm trial}(z_1,\ldots,z_N)
 =\left[\prod_{i=1}^{N}G(\bm r_i)\right]
 \prod_{i<j}(z_i-z_j)^2
 e^{-\sum_i|z_i|^2/(4\ell^2)}.
 \label{eq:KL_trial_wavefunction}
\end{equation}
Here $z_i=x_i+iy_i$ and
$G(\bm r)=(-1)^{(m+1)(n+1)}=-\eta_{\bm r}$ is the lattice sign. In the occupation basis defined by the transformed boson operators,
the same state's amplitudes are multiplied by $\prod_i\eta_{\bm r_i}$.
Since $\prod_i\eta_{\bm r_i}G(\bm r_i)=(-1)^N$, they take the Laughlin
form without site factors, up to an overall phase at fixed $N$.

\paragraph{Single-particle band.}
The site transformation leaves the one-particle spectrum unchanged.
In the positive-hopping gauge,
\begin{equation}
    h_{\rm KL}(\bm k)=2t_0\sum_{\mu=1}^{3}
    \cos(\bm k\cdot\bm d_\mu). \label{eq:KL_dispersive_band}
\end{equation}
The cell-periodic eigenvector can be chosen as $u_{\bm k}=1$
so the band has zero Berry curvature and Chern number and does not
provide the exact lattice LLL used in the main text. The original
Heisenberg model instead has three-sublattice $120^\circ$ magnetic order~\cite{HuseElser1988,JolicoeurLeGuillou1989SpinWave,
SinghHuse1992ThreeSublattice,WhiteChernyshev2007NeelOrder}.

\subsection{Exact spin Hamiltonians on non-Bravais lattices} \label{app:nonbravais_spin}
We now discuss in detail how to construct spin-$1/2$ parent Hamiltonians for Kalmeyer--Laughlin type CSLs on non-Bravais lattices, as an extension of the Bravais-lattice spin Hamiltonian mentioned in the main text.
We use the Gaussian-decay model of Li and Wu~\cite{LiWu2026LatticeGLL}. On the infinite lattice, let
$\bm\tau_\alpha$ ($\alpha=1,\ldots,n_{\rm sub}$) be
$n_{\rm sub}\ge2$ basis sites, distinct modulo the Bravais lattice, with
$\bm a_1\times\bm a_2=2\pi\ell^2$, $\bm r=\bm R+\bm\tau_\alpha$, and $\eta_{m\bm a_1+n\bm a_2}=(-1)^{m+n+mn}$.
Here $m,n\in\mathbb Z$; $\bm R$ and the cell displacement $\bm D$ are integer combinations
of $\bm a_1,\bm a_2$, and $\ell$ is the auxiliary LLL magnetic length:
there is one flux quantum per Bravais cell.

Following Li and Wu~\cite{LiWu2026LatticeGLL}, choose the site modulation
\begin{equation}
\mathcal B(\bm R+\bm\tau_\alpha)
=\eta_{\bm R}e^{-i\bm R\times\bm\tau_\alpha/(2\ell^2)}\mathcal B_\alpha,
\qquad \mathcal B_\alpha:=\mathcal B(\bm\tau_\alpha),\quad
|\mathcal B_\alpha|=1.
\label{eq:NB_site_factor}
\end{equation}
Let $\psi_{j,\bm k}$ be an auxiliary orbital in LL $j$,
labeled by the ordinary Bloch momentum of its modulated lattice samples. Then
\begin{equation}
\begin{aligned}
\psi_{j,\bm k}(\bm R+\bm\tau_\alpha)
&=\eta_{\bm R}e^{i\bm R\times\bm\tau_\alpha/(2\ell^2)}
e^{i\bm k\cdot\bm R}\psi_{j,\bm k}(\bm\tau_\alpha),\\
(\mathcal B\psi_{j,\bm k})(\bm R+\bm\tau_\alpha)
&=e^{i\bm k\cdot\bm R}\mathcal B_\alpha\psi_{j,\bm k}(\bm\tau_\alpha).
\end{aligned}
\label{eq:BN_modulated_Bloch}
\end{equation}

With $|\mathcal B_\alpha|=1$, the site modulation is unitary.
The lattice-periodic one-body Hamiltonian $h^G$ below is obtained from the generalized KM Hamiltonian~\cite{DongMueller2020ExactFlatBands} by this modulation and an overall sign reversal, giving $h^G\ge0$ (Supplemental Material, Sec.~IV.2 of
Ref.~\cite{LiWu2026LatticeGLL}).
Its hopping matrix elements,
$J^G_{\alpha\beta}(\bm D)
:=\langle\bm R+\bm D+\bm\tau_\alpha|
h^G|\bm R+\bm\tau_\beta\rangle$,
are~\cite{LiWu2026LatticeGLL}
\begin{equation}
\begin{aligned}
J^G_{\alpha\beta}(\bm D)=J_G\eta_{\bm D}\Bigl[{}
&\delta_{\alpha\beta}e^{-|\bm D|^2/(4\ell^2)}
\sum_{\gamma=1}^{n_{\rm sub}}e^{i\bm\tau_\gamma\times\bm D/\ell^2}
\\[-1pt]
&-\mathcal B_\alpha\mathcal B_\beta^*
e^{\frac{i}{2\ell^2}[(\bm\tau_\alpha+\bm\tau_\beta)\times\bm D+
\bm\tau_\beta\times\bm\tau_\alpha]}
e^{-|\bm D+\bm\tau_\alpha-\bm\tau_\beta|^2/(4\ell^2)}\Bigr].
\end{aligned}
\label{eq:NB_Gaussian_remainder}
\end{equation}
Here $J_G>0$ sets the energy scale, $\delta_{\alpha\beta}$ is the
Kronecker delta, and $\gamma$ sums over basis sites;
$\mathcal B_\alpha$ is defined in Eq.~\eqref{eq:NB_site_factor}.
The lattice states with amplitudes $\mathcal B(\bm r)\psi_{0,\bm k}(\bm r)$ are zero modes of $h^G$~\cite{LiWu2026LatticeGLL}.

Applying the hard-core boson--spin mapping gives the spin-$1/2$ parent Hamiltonian for the CSL on a non-Bravais lattice:
\begin{equation}
    H_{\rm XY}^{G} = \sum_{\bm R,\bm D}\sum_{\alpha,\beta=1}^{n_{\rm sub}} J^G_{\alpha\beta}(\bm D) S^+_{\bm R+\bm D+\bm\tau_\alpha}S^-_{\bm R+\bm\tau_\beta}. \label{eq:NB_spin_parent}
\end{equation}
The directed sum includes the uniform onsite term
$J_G(n_{\rm sub}-1)\sum_{\bm r}(S_{\bm r}^z+1/2)$,
obtained from Eq.~\eqref{eq:NB_Gaussian_remainder} at $\bm D=0$ and
$\alpha=\beta$, with $|\mathcal B_\alpha|=1$. The exact Kalmeyer--Laughlin state follows from Eq.~\eqref{eq:Psi_KL_spin} by replacing
$\eta(z_i)$ with $\mathcal B(\bm r_i)$ and using magnetic length $\ell$. Its single-particle dependence lies in the one-body kernel, and its coincidence zeros enforce the hard-core constraint, giving an exact
zero-energy ground state in the corresponding magnetization sector.

\subsection{Gauge conventions and magnetic translations}
\label{app:KM_gauge}
We explain how the site-gauge transformation absorbs
the $Z_2$ sign factor into the KM hopping phase used in the main text,
and specify the corresponding torus boundary conditions. We then
discuss magnetic translations and winding contributions to onsite
terms on finite tori.

\paragraph{Site gauges.}
To obtain the KM hopping phase used in the main text, apply the
site-sign transformation
$c^{(\eta)}_{\bm r}:=\eta_{\bm r}c_{\bm r}$,
with $\eta_{\bm r}$ defined in Sec.~\ref{app:KL_review}.
The identity
$\eta_{\bm r+\bm d}\eta_{\bm r}\eta_{\bm d}
=e^{i\bm r\times\bm d/(2\ell^2)}$
then gives, for the standard-gauge KM hopping
$t_{\bm r,\bm d}=W_{\bm d}\eta_{\bm d}
e^{i\phi\bm r\times\bm d/(2\ell^2)}$,
\begin{equation}
t^{(\eta)}_{\bm r,\bm d}=\eta_{\bm r+\bm d}\eta_{\bm r}t_{\bm r,\bm d}=W_{\bm
d}e^{i(\phi+1)\bm r\times\bm d/(2\ell^2)}.
\label{eq:KM_eta_phase}
\end{equation}
The $N$-particle wavefunction acquires $\prod_{i=1}^N\eta_{\bm r_i}$, as in
Eq.~\eqref{eq:Psi_KL_spin}. These onsite transformations preserve loop phases.

For the two-sublattice Gaussian-decay model,
choose $\bm\tau_A=-\bm\tau_B$, $\mathcal B_A=1$, and $\mathcal B_B=-1$
in Eq.~\eqref{eq:NB_Gaussian_remainder}%
~\cite{LiWu2026LatticeGLL}, and
define the site-gauge transformation
$U_G|\bm R+\bm\tau_\alpha\rangle
=\eta_{\bm R}e^{i\bm R\times\bm\tau_\alpha/(2\ell^2)}
|\bm R+\bm\tau_\alpha\rangle$.
Using the site-sign identity gives
\begin{equation}
\langle\bm r+\bm d|U_G^\dagger h^G U_G|\bm r\rangle
=J_Ge^{-|\bm d|^2/(4\ell^2)}
e^{-i\bm r\times\bm d/(2\ell^2)}.
\label{eq:GF_uniform_hopping}
\end{equation}
The phase factor in this hopping matrix element
corresponds to a uniform magnetic field in symmetric gauge.

\paragraph{Magnetic translations and torus boundary conditions.}
Let $\bm a_\mu$ denote the primitive Bravais-lattice vectors and $\phi$ the
flux per primitive cell in units of the flux quantum, with $\bm a_1\times\bm
a_2=2\pi\ell^2$.
In standard symmetric gauge, magnetic translation by a lattice vector $\bm b$
acts as $\mathsf t_{\bm b}|\bm r\rangle=e^{i\phi\bm b\times\bm
r/(2\ell^2)}|\bm r+\bm b\rangle$.
On an infinite Bravais lattice with one orbital per primitive cell, consider a hopping Hamiltonian $h$ invariant under magnetic translations by both primitive vectors, i.e., $[h,\mathsf t_{\bm a_\mu}]=0$ for $\mu=1,2$. For any such Hamiltonian, including the KM model, the hopping $J^\infty_{\bm r,\bm d}$ from $\bm r$ to $\bm r+\bm d$ obeys
\begin{equation}
J^\infty_{\bm r+\bm b,\bm d}=e^{i\phi\bm b\times\bm d/(2\ell^2)}J^\infty_{\bm
r,\bm d},
\label{eq:app_hopping_covariance}
\end{equation}
so its magnitude depends only on $\bm d$ and its onsite coefficient is
uniform.

On a torus with periods $\bm L_\mu=L_\mu\bm a_\mu$, with positive integers
$L_\mu$, impose the magnetic boundary conditions
\cite{KapitMueller2010,Shen2025,CruiseSeidel2026}
\begin{equation}
\begin{aligned}
\psi^{\rm std}(\bm r+\bm L_\mu)
&=g_\mu^{\rm std}(\bm r)\psi^{\rm std}(\bm r),\\
g_\mu^{\rm std}(\bm r)
&=e^{i\vartheta_\mu}e^{i\phi\bm L_\mu\times\bm r/(2\ell^2)},
\qquad \mu=1,2.
\end{aligned}
\label{eq:KM_torus_boundary_std}
\end{equation}
The twists $\vartheta_\mu$ are fixed modulo $2\pi$, and $N_\phi=\phi
L_1L_2\in\mathbb Z$.
In the site gauge $\psi^{(\eta)}(\bm r)=\eta_{\bm r}\psi^{\rm std}(\bm r)$,
the boundary factors become
\begin{equation}
\begin{aligned}
g_\mu^{(\eta)}(\bm r)
&=\frac{\eta_{\bm r+\bm L_\mu}}{\eta_{\bm r}}g_\mu^{\rm std}(\bm r)=e^{i(\vartheta_\mu+\pi L_\mu)}
e^{i(\phi+1)\bm L_\mu\times\bm r/(2\ell^2)}.
\end{aligned}
\label{eq:KM_torus_boundary_eta}
\end{equation}
A bulk magnetic translation $\bm b$ preserves the fixed twists precisely when
\begin{equation}
e^{i\phi\bm b\times\bm L_\mu/\ell^2}=1,\qquad \mu=1,2.
\label{eq:KM_torus_compatible_translation}
\end{equation}
Both primitive magnetic translations survive precisely when $\phi L_1,\phi
L_2\in\mathbb Z$.

\paragraph{Periodization and onsite coefficients.}
Using the standard-gauge boundary factors, choose $\bm r$ and $\bm r+\bm d$ in
one torus cell. With $g_{\bm L}$ defined by $\psi(\bm r+\bm L)=g_{\bm L}(\bm
r)\psi(\bm r)$, the torus hopping $J^T_{\bm r,\bm d}$ is
\begin{equation}
J^T_{\bm r,\bm d}=\sum_{\bm L}J^\infty_{\bm r+\bm L,\bm d-\bm L}g_{\bm L}(\bm
r),
\label{eq:CA_torus_images}
\end{equation}
where $\bm L$ runs over integer combinations of the torus periods.
Combining the hopping covariance with the boundary factors, a bulk magnetic
translation $\bm b$ multiplies each summand by $e^{i\phi\bm b\times\bm
d/(2\ell^2)}e^{-i\phi\bm b\times\bm L/\ell^2}$.
The second factor is unity under
Eq.~\eqref{eq:KM_torus_compatible_translation}, so $J^T$ obeys
Eq.~\eqref{eq:app_hopping_covariance} with translated sites identified by the
magnetic boundary conditions.

Setting $\bm d=\bm0$ gives the onsite coefficient $\epsilon_{\bm r}=J^T_{\bm
r,\bm0}$: the $\bm L=\bm0$ term is the bulk onsite coefficient, and the
remaining terms are winding contributions.
For models with this symmetry, $\phi L_1,\phi L_2\in\mathbb Z$ guarantees uniform
$\epsilon_{\bm r}$. Otherwise the winding terms can depend on the site; they
vanish for exponentially or Gaussian-decaying hopping as the periods increase
at fixed model parameters and shape.
Any bulk onsite variation remains. These coefficients enter the Zeeman
fields of Sec.~\ref{app:TR_symmetry}.

\section{Hubbard realizations and time-reversal symmetry}\label{app:hubbard_symmetry}
We extend the main-text superexchange construction to prescribed XY exchange couplings with or without an additional Ising interaction in the singly occupied Mott regime. Sections~\ref{app:fermi_hubbard} and~\ref{app:bose_hubbard} show how to choose the hopping and interaction parameters in Fermi-- and Bose--Hubbard models, respectively, to obtain these spin couplings. Section~\ref{app:TR_symmetry} shows how a uniform Zeeman field lifts the degeneracy between time-reversed CSL states with opposite magnetizations and chiralities.

\paragraph{Spin Hamiltonian.}
We match the magnitude and phase of each offsite XY exchange in Eq.~\eqref{eq:Hspin_general_nLL}. For a model built from $n_{\rm LL}$ sampled Landau levels, if the exchange magnitude depends only on the displacement $\bm d$, the offsite spin Hamiltonian takes the form
\begin{equation}
H_{{\rm XY},{\rm off}}^{(n_{\rm LL})}
=\sum_{\bm r,\bm d\ne\bm0}
W_{\bm d}^{(n_{\rm LL})}e^{i\varphi_{\bm r,\bm d}^{(n_{\rm LL})}}
S_{\bm r+\bm d}^+S_{\bm r}^-.
\label{eq:app_target_spin_parent}
\end{equation}
For site-dependent couplings, including those of general positive-weight
models and finite tori, the superexchange formulas apply bond by bond on both Bravais and non-Bravais lattices,
using $W_{\bm r,\bm d}^{(n_{\rm LL})}=|J_{\bm r,\bm d}^{(n_{\rm LL})}|$.
Reverse links have equal magnitudes and opposite phases.

\subsection{Fermi--Hubbard realization}\label{app:fermi_hubbard}
We derive the conditions under which the Fermi--Hubbard model reproduces the target spin couplings at second order and analyze the corrections.

\paragraph{Microscopic model.}
On a Bravais lattice, we use $H=H_t+H_V$, with
\begin{align}
 H_t&=-\sum_{\bm r,\bm d\ne\bm0,\sigma}
 t_{\bm d}e^{is_\sigma\vartheta_{\bm r,\bm d}}
 f_{\bm r+\bm d,\sigma}^{\dagger}f_{\bm r,\sigma},
 \label{eq:app_fermion_Ht_p}\\
 H_V&=U_{\uparrow\downarrow}\sum_{\bm r}
 n_{\bm r,\uparrow}n_{\bm r,\downarrow}
 +\frac12\sum_{\bm r,\bm d\ne\bm0,\sigma,\sigma'}
 V_{\bm d}^{\sigma\sigma'}n_{\bm r+\bm d,\sigma}n_{\bm r,\sigma'}.
\end{align}
Here $f_{\bm r,\sigma}$ are canonical fermion operators,
$n_{\bm r,\sigma}=f_{\bm r,\sigma}^{\dagger}f_{\bm r,\sigma}$,
and $s_{\uparrow,\downarrow}=\pm1$.
The sums include both bond orientations, with real
$t_{-\bm d}=t_{\bm d}$ and
$\vartheta_{\bm r+\bm d,-\bm d}=-\vartheta_{\bm r,\bm d}$.
There is no additional spin-independent orbital Peierls phase.
We take $V_{\bm d}^{\uparrow\uparrow}=V_{\bm d}^{\downarrow\downarrow}
\equiv V_{\bm d}^{s}$ and
$V_{\bm d}^{\uparrow\downarrow}=V_{\bm d}^{\downarrow\uparrow}
\equiv V_{\bm d}^{o}$, with $V_{-\bm d}^{s,o}=V_{\bm d}^{s,o}$.
Let $P_0$ project onto the singly occupied subspace,
$n_{\bm r,\uparrow}+n_{\bm r,\downarrow}=1$, where
$S_{\bm r}^+=P_0f_{\bm r,\uparrow}^\dagger f_{\bm r,\downarrow}P_0$
and $n_{\bm r,\uparrow/\downarrow}=1/2\pm S_{\bm r}^z$.
We assume a stable Mott regime with the hopping-induced kinetic-energy
scale much smaller than the charge-excitation energy.

\paragraph{XY and Ising couplings.}
Using the virtual excitation energy on this bond $\Delta_{\bm d}=U_{\uparrow\downarrow}-V_{\bm d}^o$,
second-order Schrieffer--Wolff perturbation theory gives the spin-flip
amplitude $2|t_{\bm d}|^2e^{i2\vartheta_{\bm r,\bm d}}/\Delta_{\bm d}$
and the Ising coupling $4|t_{\bm d}|^2/\Delta_{\bm d}$ per undirected
bond~\cite{BravyiDiVincenzoLoss2011}.
The density interaction projects to
\begin{equation}
 P_0H_{V,\bm r,\bm d}P_0
 =2(V_{\bm d}^s-V_{\bm d}^o)S_{\bm r+\bm d}^zS_{\bm r}^z
 +\mathrm{const.}
 \label{eq:shared_density_Ising_projection}
\end{equation}
Adding these contributions and matching the target transverse
coefficient $W_{\bm d}e^{i\varphi_{\bm r,\bm d}}$ and Ising coefficient
$\mathcal V_{\bm d}^{(2)}=\mathcal V_{-\bm d}^{(2)}$ per undirected bond gives
\begin{equation}
 \begin{gathered}
 2\vartheta_{\bm r,\bm d}=\varphi_{\bm r,\bm d}\pmod{2\pi},
 \qquad\frac{2|t_{\bm d}|^2}{\Delta_{\bm d}}=W_{\bm d},\\
 2(V_{\bm d}^s-V_{\bm d}^o)+\frac{4|t_{\bm d}|^2}{\Delta_{\bm d}}
 =\mathcal V_{\bm d}^{(2)}.
 \end{gathered}
 \label{eq:fermi_positive_Ising_matching}
\end{equation}
For the XY model, one sets $\mathcal V_{\bm d}^{(2)}=0$ on every bond. For the two-sublattice Gaussian-decay model, the phase in Eq.~\eqref{eq:GF_uniform_hopping} is reproduced by hopping phases $\mp(\bm r\times\bm d)/(4\ell^2)$ for the two internal states on each offsite bond. These phases correspond to opposite uniform effective magnetic fields. For the deformed Laughlin models, take $\mathcal V_{\pm\bm\delta_\lambda}^{(2)}=\mathcal V_{\bm\delta_\lambda}^{(2)}>0$ on the selected pair displacements and zero otherwise. Then
\begin{equation}
 \frac12\sum_{\bm r,\bm d\ne\bm0}\mathcal V_{\bm d}^{(2)}
 S_{\bm r+\bm d}^zS_{\bm r}^z
 =\sum_{\bm r,\lambda}\mathcal V_{\bm\delta_\lambda}^{(2)}
 S_{\bm r+\bm\delta_\lambda}^zS_{\bm r}^z.
 \label{eq:target_Ising_link_counting}
\end{equation}
The linear terms from the target density products are supplied by the
Zeeman fields discussed in Sec.~\ref{app:TR_symmetry}.

\paragraph{Validity and higher-order corrections.}
With $A_{\bm d}:=V_{\bm d}^s-V_{\bm d}^o$, the virtual excitation energy
for a hop of spin $\sigma$ from $\bm r$ to $\bm r+\bm d$ is
$\Delta_{\bm d}+s_\sigma X_{\bm r,\bm d}$, where
\begin{equation}
 X_{\bm r,\bm d}:=
 \sum_{\bm r'\ne\bm r,\bm r+\bm d}
 \left(A_{\bm r+\bm d-\bm r'}-A_{\bm r-\bm r'}\right)S_{\bm r'}^z.
 \label{eq:CA_SW_environment}
\end{equation}
Neglecting the contribution from the surrounding spins requires $\Delta_{\bm d}>0$ and
$\|X_{\bm r,\bm d}\|/\Delta_{\bm d}\ll1$ on each bond.
In the Hermitian second-order expression, the spin-flip amplitude
contains $(\Delta_{\bm d}+X_{\bm r,\bm d})^{-1}
+(\Delta_{\bm d}-X_{\bm r,\bm d})^{-1}$, so its linear correction cancels.
The diagonal processes give
\begin{equation}
 \delta H_{\bm r,\bm d,\rm ex}^{(2)}
 =\frac{2|t_{\bm d}|^2}{\Delta_{\bm d}^2}
 X_{\bm r,\bm d}(S_{\bm r}^z-S_{\bm r+\bm d}^z)
 +O\!\left(\frac{|t_{\bm d}|^2\|X_{\bm r,\bm d}\|^2}{\Delta_{\bm d}^3}\right),
 \label{eq:CA_SW_environment_correction}
\end{equation}
whose linear term contains only two-spin Ising interactions.
Let $t$ and $\Delta$ denote characteristic hopping and charge-excitation
energy scales. In the strong-coupling Mott regime, the target Ising coupling can be much larger than $t^2/\Delta$ provided $\|X_{\bm r,\bm d}\|\ll\Delta_{\bm d}$ on each bond. For the pure XY model, $\mathcal V_{\bm d}^{(2)}=0$ on every bond. In this case, or more generally when $\mathcal V_{\bm d}^{(2)}=O(t^2/\Delta)$, Eq.~\eqref{eq:fermi_positive_Ising_matching} gives $A_{\bm d}=O(t^2/\Delta)$. For the decaying interactions considered here at fixed geometry, the correction in Eq.~\eqref{eq:CA_SW_environment_correction} is then $O(t^4/\Delta^3)$. For the equal-magnitude, opposite-phase hopping and spin-symmetric density interactions used here, the half-filled model has no cubic hopping contribution~\cite{Garuchava2025SpinDependentFlux,SenChitra1995}. The leading corrections in this regime are therefore $O(t^4/\Delta^3)$.

\subsection{Bose--Hubbard realization}
\label{app:bose_hubbard}
We use the hopping to set the XY exchange, as in the main text, and an additional offsite density interaction to tune the Ising coupling. For $\bm d\in\mathcal D_+$, matching the Bose--Hubbard exchange in Eq.~\eqref{eq:H_exchange_general} to the target XY coupling in Eq.~\eqref{eq:app_target_spin_parent} gives
\begin{equation}
    |t_{\bm d}^{(n_{\rm LL})}| =
    \sqrt{\frac{U_{\uparrow\downarrow}}{2}W_{\bm d}^{(n_{\rm LL})}},
    \qquad
    2\theta_{\bm r,\bm d}^{(n_{\rm LL})}
    = \varphi_{\bm r,\bm d}^{(n_{\rm LL})}+\pi
    \pmod{2\pi}.
    \label{eq:app_boson_square_root_p}
\end{equation}

On each undirected bond, the projected density interaction [Eq.~\eqref{eq:shared_density_Ising_projection}] adds to the bosonic Ising exchange $J_{\bm d}^{z,B}$, giving
\begin{equation}
    2(V_{\bm d}^{s}-V_{\bm d}^{o})+J_{\bm d}^{z,B}
    = \mathcal V_{\bm d}^{(2)}.
    \label{eq:boson_positive_Ising_matching}
\end{equation}
For $U_\uparrow=U_\downarrow=2U_{\uparrow\downarrow}$, this reduces to $2(V_{\bm d}^{s}-V_{\bm d}^{o})=\mathcal V_{\bm d}^{(2)}$.

We take $V^{s,o}=O(t^2/U)$, where $t$ and $U$ are characteristic hopping and onsite-interaction scales. Their contributions to the virtual excitation energies change the superexchange at $O(t^4/U^3)$. Three-hop processes around triangular loops generally give corrections of order $t^3/U^2$ to both the transverse and Ising exchange~\cite{PachosPlenio2004}.

\subsection{Time-reversal symmetry and Zeeman fields}\label{app:TR_symmetry}
We examine how spin time reversal relates CSL states with opposite magnetizations and chiralities, and how Zeeman fields split their energies. We also examine time-reversal symmetry in the Hubbard realizations.

\paragraph{Spin time reversal and CSL partners.}
For an offsite bond $\bm r\to\bm r+\bm d$,
$J_{\bm r+\bm d,-\bm d}=J_{\bm r,\bm d}^{*}$.
In the spin axes fixed by Eq.~\eqref{eq:MM_map}, the bond Hamiltonian is
\begin{equation}
\begin{aligned}
 H_{\bm r,\bm d}
 &=J_{\bm r,\bm d}S_{\bm r+\bm d}^+S_{\bm r}^-
 +J_{\bm r,\bm d}^*S_{\bm r}^+S_{\bm r+\bm d}^-\\
 &=2\operatorname{Re}J_{\bm r,\bm d}
 (S_{\bm r+\bm d}^xS_{\bm r}^x+S_{\bm r+\bm d}^yS_{\bm r}^y)
 +2\operatorname{Im}J_{\bm r,\bm d}\,\kappa_{\bm r,\bm d}^z.
\end{aligned}
\label{eq:TR_XY_DM_bond}
\end{equation}
Here $\kappa_{\bm r,\bm d}^z:=(\bm S_{\bm r+\bm d}\times\bm S_{\bm r})_z$,
and $2\operatorname{Im}J_{\bm r,\bm d}$ is the Dzyaloshinskii--Moriya (DM) coefficient for the
chosen orientation. Summing $H_{\bm r,\bm d}$ once per undirected bond
gives $H_{\rm XY}^{\rm off}$.

In the $S^z$ product basis, let $\mathcal K$ conjugate coefficients.
With $\hbar=1$, physical spin time reversal obeys
\begin{equation}
 \mathcal T=e^{-i\pi\sum_{\bm r}S_{\bm r}^y}\mathcal K,
 \qquad \mathcal T\bm S_{\bm r}\mathcal T^{-1}=-\bm S_{\bm r},
 \qquad \mathcal T\kappa_{\bm r,\bm d}^z\mathcal T^{-1}
 =\kappa_{\bm r,\bm d}^z.
\label{eq:TR_definition_actions}
\end{equation}
Both the XY and DM terms are therefore time-reversal invariant.
A CSL eigenstate of $H_{\rm XY}^{\rm off}$ in the $S_{\rm tot}^z=M$ sector
has a time-reversed partner with the same energy, magnetization $-M$,
and opposite chirality. A uniform field $-hS_{\rm tot}^z$ shifts their
energies by $-hM$ and $+hM$, respectively.

\paragraph{Effective Zeeman fields.}
The onsite hopping coefficient $\epsilon_{\bm r}:=J_{\bm r,\bm0}$ contributes a Zeeman term under the hard-core boson--spin map:
\begin{equation}
H_{\rm XY}=H_{\rm XY}^{\rm off}+\sum_{\bm r}\epsilon_{\bm r}(S_{\bm r}^z+1/2).
\label{eq:CA_full_spin_parent}
\end{equation}
Adding the external field $-\sum_{\bm r}h_{\bm r}^{\rm ext}S_{\bm r}^z$ gives $H(\{h_{\bm r}\})=H_{\rm XY}^{\rm off}-\sum_{\bm r}h_{\bm r}S_{\bm r}^z$, up to a constant, with $h_{\bm r}=h_{\bm r}^{\rm ext}-\epsilon_{\bm r}$. Thus $\mathcal T H(\{h_{\bm r}\})\mathcal T^{-1}=H(\{-h_{\bm r}\})$.

To separate the uniform field from onsite variations, write $\epsilon_{\bm r}=\bar\epsilon+\delta\epsilon_{\bm r}$, where $\bar\epsilon=N_s^{-1}\sum_{\bm r}\epsilon_{\bm r}$ ($\bar\epsilon=W_0$ for the infinite-lattice KM model). For the uniform external field $h_{\rm ext}=\bar\epsilon+h$ used in Fig.~\ref{fig:kl-numerics}(e),
\begin{equation}
H(h,\{\delta\epsilon_{\bm r}\})=H_{\rm XY}^{\rm off} +\sum_{\bm r}\delta\epsilon_{\bm r}S_{\bm r}^z-hS_{\rm tot}^z,\qquad \mathcal T H(h,\{\delta\epsilon_{\bm r}\})\mathcal T^{-1} =H(-h,\{-\delta\epsilon_{\bm r}\}).
\label{eq:CA_plateau_hamiltonian}
\end{equation}
At fixed $\{\delta\epsilon_{\bm r}\}$, let $E_{\min}(M)$ be the lowest energy at $h=0$ in the $S_{\rm tot}^z=M$ sector. A magnetization plateau occurs when the same sector minimizes $E_{\min}(M)-hM$ over a finite field interval.

For the deformed Laughlin sequence, the mapping $n_i=S_i^z+1/2$ gives a linear term $(S_i^z+S_j^z)/2$ in each pair-density product. Summing over the translation-invariant bonds of the torus gives
\begin{equation}
H_{\rm XY}+H_{2,\delta}^{(n_2)} = H_{\rm XY}^{\rm off} +\sum_{\bm r,\lambda}\mathcal V_{\bm\delta_\lambda}^{(2)} S_{\bm r}^zS_{\bm r+\bm\delta_\lambda}^z +\sum_{\bm r}\delta\epsilon_{\bm r}S_{\bm r}^z +\left(\bar\epsilon+\sum_\lambda\mathcal V_{\bm\delta_\lambda}^{(2)}\right) S^z_{\rm tot}+\mathrm{const.}
\label{eq:TR_Laughlin_density_expansion}
\end{equation}
The Ising exchange is invariant under spin time reversal. Adding the external field $-h_{\rm ext}S_{\rm tot}^z$ gives the net uniform field
\begin{equation}
h_{\rm net}=h_{\rm ext}-\bar\epsilon-\sum_\lambda\mathcal V_{\bm\delta_\lambda}^{(2)}.
\label{eq:TR_Laughlin_net_field}
\end{equation}
When all sites have the same onsite energy, zero applied Zeeman field in the Hubbard model corresponds to $h_{\rm net}=0$ in the spin model, reached at $h_{\rm ext}=\bar\epsilon+\sum_\lambda\mathcal V_{\bm\delta_\lambda}^{(2)}$. At $h_{\rm net}=0$, any spatial variation in the onsite energies leaves a local Zeeman term $\sum_{\bm r}\delta\epsilon_{\bm r}S_{\bm r}^{z}$. For the higher-spin models ($k>1$) in Sec.~\ref{app:spin_truncated_boson} and the displaced-density models with higher-body interactions ($k\ge2$) in Sec.~\ref{app:positive_delta_interactions}, time reversal generally also changes the interaction terms.

\paragraph{Hubbard realizations.}
Under the map in Eq.~\eqref{eq:MM_map}, spin time reversal sends $b_i$ to $-b_i^\dagger$ and $n_i$ to $1-n_i$. In the singly occupied Hubbard subspace, $N_b=\sum_i n_{i,\uparrow}$, so exchanging the two internal states sends $N_b$ to $N_s-N_b$ at fixed total atom number. For $\bm q_i=(q_{i,\uparrow},q_{i,\downarrow})^T$, with $q=a$ for bosons and $q=f$ for fermions, define the antiunitary operation on the full Hubbard Fock space by
\begin{equation}
\mathcal T_H\bm q_i\mathcal T_H^{-1}=G\bm q_i,\qquad G:=i\sigma_y,\qquad \mathcal T_H i\mathcal T_H^{-1}=-i.
\label{eq:TR_Hubbard_doublet_action}
\end{equation}
It reverses $\bm S_i$ in the singly occupied subspace and represents physical time reversal when the chosen internal states transform as specified above. On each bond, the two components have opposite Peierls phases, so time reversal maps the hopping term for one component to that for the other and leaves their sum invariant. With the density interactions of Secs.~\ref{app:fermi_hubbard} and~\ref{app:bose_hubbard} and $U_\uparrow=U_\downarrow$ for bosons, both Hubbard Hamiltonians are therefore invariant under $\mathcal T_H$ at zero applied field.\phantomsection\label{trreview:TR07}

A uniform energy splitting $h$ between the two internal states gives the Zeeman field term $H_Z=-(h/2)\sum_i(n_{i,\uparrow}-n_{i,\downarrow})$. This field shifts all states with magnetization $M$ by $-hM$. Since a virtual hop preserves $M$, its energy cost is independent of the field.

\section{Generalized spin Hamiltonians and exact states} \label{app:generalizations}
This appendix develops the three extensions beyond the KL CSL introduced in the main text. Section~\ref{app:projector_mll} gives the multi-LL and higher-Chern projector constructions, Sec.~\ref{app:positive_delta_interactions} constructs deformed Laughlin and dressed Moore--Read zero modes, and Sec.~\ref{app:spin_truncated_boson} derives the higher-spin mapping for Read--Rezayi states.

\subsection{Multi-LL and higher-Chern constructions} \label{app:projector_mll}
We give the two projector constructions used in Route~1. For the multi-LL parton models, we choose the hopping so that the lowest $n_{\rm LL}=m+n-1$ Landau levels, sampled on the lattice, have zero energy. With the two effective fields and torus boundary phases chosen as described below, this space supports the parton state $\chi_m\chi_n$. We extend the explicit magnetic Bloch construction of Cruise and Seidel~\cite{CruiseSeidel2026} from flux $\phi=1/q$ to $\phi=p/q$ and determine the Chern number carried by each sampled LL. The exact spin states lie in sectors with $m_0=2\phi\nu_b-1$, which are candidates for magnetization plateaus. For the Halperin-type states, we use the non-Bravais construction of Li and Wu~\cite{LiWu2026LatticeGLL} to obtain a spin Hamiltonian from a single ideal higher-Chern band.

Let $\Lambda=\{r=u a_1+v a_2:u,v\in\mathbb Z\}$, with $a_1,a_2\in\mathbb C$ and corresponding real vectors $\bm a_1,\bm a_2$.
We use $\bm a_1\times\bm a_2=\operatorname{Im}(a_1^*a_2)=2\pi\ell^2>0$.
The flux per primitive cell is $\phi=p/q$, where $0<p<q$ are coprime
integers; thus $\ell_B^2=q\ell^2/p$. Choose positive integers
$\kappa_1\kappa_2=q$ and magnetic periods
$\tilde a_1=\kappa_1a_1$, $\tilde a_2=\kappa_2a_2$. Their cell contains $q$ sites and $p$ flux quanta, with
$\tilde{\bm a}_1\times\tilde{\bm a}_2=2\pi q\ell^2=2\pi p\ell_B^2$.

The two magnetic-cell translations commute. We label their common eigenstates by
\begin{equation}
\mathsf t_{\tilde{\bm a}_\mu}\Psi_{\bm k}
=e^{-i\bm k\cdot\tilde{\bm a}_\mu}\Psi_{\bm k},\qquad \mu=1,2.
\label{eq:BN_Bloch_eigenvalue}
\end{equation}
At fixed magnetic Bloch momentum $\bm k$, each Landau level contains $p$ independent orbitals, labeled by $s=0,\ldots,p-1$. These orbitals have the same eigenvalues under both magnetic-cell translations.

\paragraph{Magnetic Bloch states at rational flux.}
\begingroup
The theta-function orbitals below give the amplitudes that we sample on the lattice. They also determine how the orbitals are related at opposite edges of the magnetic Brillouin zone. Write $\tau=\tilde a_2/\tilde a_1$ with $\operatorname{Im}\tau>0$. In the translation convention of Eq.~\eqref{eq:BN_Bloch_eigenvalue}, set
\[
 \eta_{\bm k}=\bm k\cdot\tilde{\bm a}_1/(2\pi),\qquad
 \xi_{\bm k}=-\bm k\cdot\tilde{\bm a}_2/(2\pi).
\]
We use the theta function
\begin{equation}
    \vartheta_{\eta,\xi}(u|\tau)
    :=\sum_{l\in\mathbb Z}
    e^{i\pi(l+\eta)^2\tau+i2\pi(l+\eta)(u+\xi)}.
    \label{eq:rf_theta}
\end{equation}
In symmetric gauge $\bm A_S=(-y,x)/(2\ell_B^2)$, the LLL orbital with labels $(\bm k,s)$ is
\begin{equation}
    \Psi^{S,(0)}_{\bm k;s}
    =F_{\bm k;s}(z)e^{-|z|^2/(4\ell_B^2)},\qquad
    F_{\bm k;s}(z)=\mathcal N_p
    e^{\tilde a_1^*z^2/(4\ell_B^2\tilde a_1)}
    \vartheta_{(\eta_{\bm k}+s)/p,\xi_{\bm k}}
    \!\left(\frac{pz}{\tilde a_1}\middle|p\tau\right).
    \label{eq:rf_symmetric_basis}
\end{equation}
We fix continuum normalization by
$\int_{\rm cell}d^2r\,|\Psi|^2/(2\pi\ell^2)=1$, for which
$\mathcal N_p^2=2\pi\ell^2/(|\tilde a_1|\sqrt{\pi\ell_B^2})$.

In Eq.~\eqref{eq:rf_symmetric_basis}, shifting $z$ by $\tilde a_1$ shifts the theta-function argument by $p$ and gives the factor $e^{2\pi i(\eta_{\bm k}+s)}=e^{2\pi i\eta_{\bm k}}$. Shifting $z$ by $\tilde a_2$ shifts the argument by $p\tau$ and gives $e^{-i\pi p\tau-2\pi i(pz/\tilde a_1+\xi_{\bm k})}$. Combining these shifts with the gauge factor in the magnetic translations gives
\begin{equation}
    \mathsf t_{\tilde{\bm a}_1}\Psi^{S,(0)}_{\bm k;s}
    =e^{-i2\pi\eta_{\bm k}}\Psi^{S,(0)}_{\bm k;s},\qquad
    \mathsf t_{\tilde{\bm a}_2}\Psi^{S,(0)}_{\bm k;s}
    =e^{i2\pi\xi_{\bm k}}\Psi^{S,(0)}_{\bm k;s}.
    \label{eq:rf_translation_eigenvalues}
\end{equation}
Equivalently, the coordinate amplitudes obey
\begin{equation}
    \Psi_{\bm k}(\bm r+\tilde{\bm a}_\mu)
    =e^{i\tilde{\bm a}_\mu\times\bm r/(2\ell_B^2)}
    e^{i\bm k\cdot\tilde{\bm a}_\mu}\Psi_{\bm k}(\bm r),
    \qquad\mu=1,2.
    \label{eq:rf_coordinate_sewing}
\end{equation}
Here $\Psi_{\bm k}$ denotes an LL orbital in this Bloch sector. On a finite torus, the boundary twists along the full periods
$\bm L_1$ and $\bm L_2$ select the allowed values of $\bm k$, as specified below Eq.~\eqref{eq:rf_full_hopping}.

The unnormalized ladder operators are
$\hat a_S=-\partial_{\bar z}-z/(4\ell_B^2)$ and
$\hat a_S^\dagger=\partial_z-\bar z/(4\ell_B^2)$, with
$[\hat a_S,\hat a_S^\dagger]=1/(2\ell_B^2)$.
We label the Landau levels by $j=0,\ldots,n_{\rm LL}-1$, with $j=0$ denoting the LLL. Applying the raising operator $j$ times gives the normalized orbital
\begin{equation}
    \Psi^{S,(j)}_{\bm k;s}
    =\frac{(\sqrt2\ell_B)^j}{\sqrt{j!}}
    (\hat a_S^\dagger)^j\Psi^{S,(0)}_{\bm k;s}
    =\frac{(\sqrt2\ell_B)^j}{\sqrt{j!}}
    e^{-|z|^2/(4\ell_B^2)}
    \sum_{l=0}^{j}\binom jl
    \left(-\frac{\bar z}{2\ell_B^2}\right)^l
    \partial_z^{j-l}F_{\bm k;s}(z).
    \label{eq:rf_ll_ladder}
\end{equation}
The raising operator preserves both magnetic-translation eigenvalues.
At $p=1$, $s=0$ and the theta functions reduce to the previous
one-flux basis; the fixed LL normalization factors do not change its
sampled subspaces.
\endgroup

\paragraph{Sampled LL blocks and composite projector.}
\begingroup
We sample the lowest $n_{\rm LL}$ continuum Landau levels on the lattice. In a reference magnetic cell, label the $q$ sites by $\iota=1,\ldots,q$. Their positions are $\tilde{\bm r}_\iota=x_\iota\bm a_1+y_\iota\bm a_2+\bm\rho_0$, where $x_\iota,y_\iota$ are integers and $\bm\rho_0$ is a common displacement of the lattice sites relative to the origin used for the continuum wavefunctions. At fixed $\bm k$, the amplitude on site $\iota$ is the value of the continuum wavefunction
$\Psi^{S,(j)}_{\bm k;s}(\tilde r_\iota,\bar{\tilde r}_\iota)$.
In the orthonormal site basis $\{|\iota\rangle\}_{\iota=1}^q$ of the magnetic cell, the corresponding magnetic Bloch state on the lattice is
\begin{equation}
 |\Psi^{S,(j)}_{\bm k;s}\rangle
 :=\sum_{\iota=1}^{q}
   \Psi^{S,(j)}_{\bm k;s}(\tilde r_\iota,\bar{\tilde r}_\iota)\,|\iota\rangle .
 \label{eq:sampled_ll_vectors}
\end{equation}
The $p$ sampled states from level $j$ form the columns of the
$q\times p$ matrix $\mathsf V_{\bm k,j}
=(|\Psi^{S,(j)}_{\bm k;0}\rangle,\ldots,|\Psi^{S,(j)}_{\bm k;p-1}\rangle)$.
The sampled wavefunctions obey the same magnetic Bloch boundary conditions, but need not be orthonormal in the lattice inner product. Combining all $n_{\rm LL}$ blocks gives the full matrix of sampled states and its Gram matrix,
\begin{equation}
 \mathsf V_{\bm k}
 =(\mathsf V_{\bm k,0},\ldots,\mathsf V_{\bm k,n_{\rm LL}-1}),
 \qquad G_{\bm k}=\mathsf V_{\bm k}^{\dagger}\mathsf V_{\bm k}.
 \label{eq:sampled_prefix_gram}
\end{equation}
We assume that all $pn_{\rm LL}$ sampled columns remain linearly independent at every $\bm k$, so $G_{\bm k}$ is positive definite.

To isolate the contribution from level $j$, we use block Gram--Schmidt orthogonalization. Let $P_{\bm k}^{[j-1]}$ project onto the span of all lower sampled levels, with $P_{\bm k}^{[-1]}=0$. The $s$th column of $\overline{\mathsf V}_{\bm k,j}$ is the state
$(1-P_{\bm k}^{[j-1]})|\Psi^{S,(j)}_{\bm k;s}\rangle$ in the site basis.
Symmetric orthonormalization of these $p$ columns gives the columns of
$\mathsf U_{\bm k,j}
=(|u_{\bm k}^{(j,0)}\rangle,\ldots,|u_{\bm k}^{(j,p-1)}\rangle)$.

For any single-particle state $|\psi\rangle$ in the same magnetic Bloch sector, $\langle u_{\bm k}^{(j,s)}|\psi\rangle$ is its amplitude along the $s$th orthonormal orbital. Summing the corresponding components gives the projection onto the part of level $j$ orthogonal to the lower levels:
\[
 \Pi_{\bm k,j}|\psi\rangle
 =\sum_{s=0}^{p-1}|u_{\bm k}^{(j,s)}\rangle
                  \langle u_{\bm k}^{(j,s)}|\psi\rangle .
\]
Explicitly,
\begin{equation}
\begin{gathered}
  \overline{\mathsf V}_{\bm k,j}
  =(1-P_{\bm k}^{[j-1]})\mathsf V_{\bm k,j},
  \qquad
  |u_{\bm k}^{(j,s)}\rangle
  =\sum_{s'=0}^{p-1}
    \Bigl[\bigl(\overline{\mathsf V}_{\bm k,j}^{\dagger}
                \overline{\mathsf V}_{\bm k,j}\bigr)^{-1/2}\Bigr]_{s's}
    (1-P_{\bm k}^{[j-1]})|\Psi^{S,(j)}_{\bm k;s'}\rangle,
  \\[3pt]
  \Pi_{\bm k,j}
  =\sum_{s=0}^{p-1}
    |u_{\bm k}^{(j,s)}\rangle\langle u_{\bm k}^{(j,s)}|
  =\mathsf U_{\bm k,j}\mathsf U_{\bm k,j}^{\dagger},
  \qquad
  P_{\bm k}^{[j]}
  =P_{\bm k}^{[j-1]}+\Pi_{\bm k,j}
  =\sum_{j'=0}^{j}\sum_{s=0}^{p-1}
    |u_{\bm k}^{(j',s)}\rangle\langle u_{\bm k}^{(j',s)}|.
\end{gathered}
\label{eq:ordered_lattice_gll}
\end{equation}
Each level thus contributes a rank-$p$ projector. At $p=1$, this construction gives the ordered lattice generalized Landau levels (GLLs)~\cite{LiuMeraFujimotoOzawaWang2025,LiWu2026LatticeGLL}. For $p>1$, unitary rotations among the $p$ orthonormal orbitals of a level leave $\Pi_{\bm k,j}$ unchanged.

Summing $\Pi_{\bm k,j}$ over $j$ gives the projector onto the zero-energy subspace of the hopping Hamiltonian,
\begin{equation}
 P_{\bm k}^{\parallel}=\sum_{j=0}^{n_{\rm LL}-1}\Pi_{\bm k,j}
 =\mathsf V_{\bm k}G_{\bm k}^{-1}\mathsf V_{\bm k}^\dagger,\qquad
 P_{\bm k}^{\perp}=1-P_{\bm k}^{\parallel}.
 \label{eq:composite_sampled_projector}
\end{equation}
The zero-energy subspace has dimension $n_{\rm flat}=pn_{\rm LL}$. We take $pn_{\rm LL}<q$ so that the complement, of dimension $n_{\rm disp}=q-pn_{\rm LL}$, is nonempty.

\paragraph{Chern number of each sampled Landau level.}
When the sampled orbitals remain linearly independent throughout the MBZ, sampling preserves the Chern number of the continuum LL bundle~\cite{AvronSeilerSimon1983Homotopy}. Here we verify explicitly that at flux $p/q$ the $p$ orbitals contributed by each Landau level together carry Chern number one. To compute this Chern number, we compare the bases of magnetic Bloch states on opposite edges of the magnetic Brillouin zone.

Parameterize the oriented MBZ by
$t_\mu=\bm k\cdot\tilde{\bm a}_\mu/(2\pi)\in[0,1]$
($\mu=1,2$), and write $\bm t=(t_1,t_2)$,
so $\eta_{\bm k}=t_1$ and $\xi_{\bm k}=-t_2$.
We abbreviate $\Psi^{S,(j)}_{\bm k(\bm t);s}$ as $\Psi^{S,(j)}_{\bm t;s}$. The following shifts of the theta characteristics determine the boundary relations:
\begin{equation}
    \vartheta_{\eta+1,\xi}(u|\tau)=\vartheta_{\eta,\xi}(u|\tau),\qquad
    \vartheta_{\eta,\xi-1}(u|\tau)=e^{-2\pi i\eta}\vartheta_{\eta,\xi}(u|\tau).
    \label{eq:rf_characteristic_shifts}
\end{equation}
The first identity follows by reindexing $l$. In the second, each summand acquires $e^{-2\pi i(l+\eta)}=e^{-2\pi i\eta}$.

Thus $t_1\mapsto t_1+1$ cycles $s\mapsto s+1\bmod p$,
while $t_2\mapsto t_2+1$ supplies a phase.
Since $\hat a_S^\dagger$ is independent of $\bm t$ and this phase
is independent of position, each of the lowest $n_{\rm LL}$ Landau levels obeys
\begin{equation}
    \Psi^{S,(j)}_{(t_1+1,t_2);s}
    =\Psi^{S,(j)}_{(t_1,t_2);(s+1)\bmod p},\qquad
    \Psi^{S,(j)}_{(t_1,t_2+1);s}
    =e^{-2\pi i(s+t_1)/p}\Psi^{S,(j)}_{(t_1,t_2);s}.
    \label{eq:rf_nth_ll_component_sewing}
\end{equation}

Write $\mathsf V_j(\bm t):=\mathsf V_{\bm k(\bm t),j}$ and
$\mathsf U_j(\bm t):=\mathsf U_{\bm k(\bm t),j}$, with the same
fixed magnetic-cell site basis as in Eq.~\eqref{eq:sampled_ll_vectors}. Sampling Eq.~\eqref{eq:rf_nth_ll_component_sewing} gives
$\mathsf V_j(1,t_2)=\mathsf V_j(0,t_2)\mathsf S$ and
$\mathsf V_j(t_1,1)=\mathsf V_j(t_1,0)\mathsf D(t_1)$, where
the $p\times p$ unitary matrices act on the orbital columns:
\begin{equation}
    \mathsf S_{ss'}=\delta_{s,(s'+1)\bmod p},\qquad
    \mathsf D_{ss'}(t_1)=\delta_{ss'}e^{-2\pi i(s+t_1)/p},\qquad
    s,s'=0,\ldots,p-1. \label{eq:rf_sewing_matrices}
\end{equation}

Let $g$ denote either boundary matrix $\mathsf S$ or $\mathsf D(t_1)$, and write $\overline{\mathsf V}_j(\bm t)=\overline{\mathsf V}_{\bm k(\bm t),j}$. Right multiplication by $g$ preserves the span of all lower levels, so $\overline{\mathsf V}_j$ transforms as $\overline{\mathsf V}_j g$. Its Gram matrix transforms by unitary conjugation. The inverse square root in Eq.~\eqref{eq:ordered_lattice_gll} transforms by the same conjugation, giving $\mathsf U_j\mapsto\mathsf U_j g$. Thus
\begin{equation}
    \mathsf U_j(1,t_2)=\mathsf U_j(0,t_2)\mathsf S,\qquad
    \mathsf U_j(t_1,1)=\mathsf U_j(t_1,0)\mathsf D(t_1).
    \label{eq:rf_block_sewing}
\end{equation}

The determinants of the two boundary matrices are
\begin{equation}
    \det\mathsf S=(-1)^{p-1},\qquad
    \det\mathsf D(t_1)=(-1)^{p-1}e^{-2\pi i t_1}.
    \label{eq:rf_sewing_determinants}
\end{equation}

Define $\partial_\mu:=\partial/\partial t_\mu$ and the trace
Berry connection
$\mathcal A_\mu^{(j)}[\mathsf U_j]
:=-i\operatorname{Tr}(\mathsf U_j^\dagger\partial_\mu\mathsf U_j)$.
Under $\mathsf U_j\mapsto\mathsf U_j g$, it changes by
$-i\operatorname{Tr}(g^\dagger\partial_\mu g)
=-i\partial_\mu\log\det g$, using $g^{-1}=g^\dagger$.
Stokes' theorem, with counterclockwise boundary orientation, gives
\begin{equation}
\begin{aligned}
C_j&=\frac1{2\pi}\int_{[0,1]^2}
(\partial_1\mathcal A_2^{(j)}-\partial_2\mathcal A_1^{(j)})
\,dt_1dt_2\\
&=\frac1{2\pi i}\left[
\int_0^1\partial_2\log\det\mathsf S\,dt_2
-\int_0^1\partial_1\log\det\mathsf D(t_1)\,dt_1\right]=1.
\end{aligned}
\label{eq:rf_chern_from_sewing}
\end{equation}

At $p=1$, $C_j$ is the Chern number of a single lattice GLL;
at $p>1$, it is the total Chern number of the $p$-dimensional
orthogonalized $j$th-LL subspace.
\endgroup

\paragraph{Hopping Hamiltonian and single-particle spectrum.}
\begingroup
We choose the hopping Hamiltonian to vanish on the sampled Landau-level orbitals and to be positive on their orthogonal complement.
Keep the site ordering $\iota=1,\ldots,q$ of Eq.~\eqref{eq:sampled_ll_vectors}
and write $|\tilde{\bm r}_{\iota}\rangle=|\iota\rangle$ for the site at
$\tilde{\bm r}_{\iota}$ in the reference cell. Order the sampled states
by $\alpha=pj+s$. For $1\le\beta_1<\cdots<\beta_{n_{\rm disp}-1}\le q$,
write $\boldsymbol\beta=(\beta_1,\ldots,\beta_{n_{\rm disp}-1})$ and define
\begin{equation}
 \langle\tilde{\bm r}_{\iota}|w_{\bm k,\boldsymbol\beta}\rangle
 :=\epsilon_{\beta_1\cdots\beta_{n_{\rm disp}-1}
                 \iota\,\iota_0\cdots\iota_{n_{\rm flat}-1}}
 \prod_{j=0}^{n_{\rm LL}-1}\prod_{s=0}^{p-1}
 \langle\Psi^{S,(j)}_{\bm k;s}|\tilde{\bm r}_{\iota_{pj+s}}\rangle .
 \label{eq:projector_complement_vectors}
\end{equation}
The Levi--Civita symbol uses the chosen site ordering, with $\epsilon_{1\cdots q}=1$.
Here $\iota$ is fixed and each $\iota_\alpha$ is summed independently
over $1,\ldots,q$. For $n_{\rm disp}=1$, the tuple $\boldsymbol\beta$ is empty.
Antisymmetry gives $\langle\Psi^{S,(j)}_{\bm k;s}|w_{\bm k,\boldsymbol\beta}\rangle=0$
for every $(j,s)$. Under the full-rank assumption these vectors span the
orthogonal complement. For positive coefficients $C_{\boldsymbol\beta}$,
the Bloch Hamiltonian is
\begin{equation}
 h_{\bm k}=\sum_{\boldsymbol\beta}C_{\boldsymbol\beta}
 |w_{\bm k,\boldsymbol\beta}\rangle\langle w_{\bm k,\boldsymbol\beta}|
 \succeq0,\qquad
 \ker h_{\bm k}=\operatorname{span}\{|\Psi^{S,(j)}_{\bm k;s}\rangle\}_{j,s}.
 \label{eq:rf_positive_block}
\end{equation}

For the spin models $H_{mn}$ in Appendix~\ref{app:numerics}, we set all coefficients $C_{\boldsymbol\beta}=1$.
The cofactor identity~\cite[Appendix~A]{CruiseSeidel2026} then gives
\[
 \sum_{\boldsymbol\beta}|w_{\bm k,\boldsymbol\beta}\rangle
 \langle w_{\bm k,\boldsymbol\beta}|=(\det G_{\bm k})P_{\bm k}^{\perp}.
\]
The resulting Hamiltonian and its spectrum are
\begin{equation}
    E_{\bm k}:=\det G_{\bm k}>0,\qquad
    h_{\bm k}=E_{\bm k}P^\perp_{\bm k},\qquad
    \operatorname{Spec}h_{\bm k}
    =\{0^{\times pn_{\rm LL}},E_{\bm k}^{\times(q-pn_{\rm LL})}\}.
    \label{eq:rf_equal_weight_bands}
\end{equation}

The superscripts denote multiplicities. The normalization of the sampled orbitals fixes $E_{\bm k}=\det G_{\bm k}$.
Choosing $h_{\bm k}=E_0P^\perp_{\bm k}$ with constant $E_0>0$ flattens the positive-energy bands without changing the kernel, but generally changes the hopping amplitudes.
For any choice of positive coefficients, the full-rank assumption and continuity of $h_{\bm k}$ on the compact MBZ ensure a nonzero single-particle gap above the zero-energy manifold.

For the Bravais construction above, write
$\bm r=\tilde{\bm r}_{\iota}+\bm L_{AB}$,
where $\tilde{\bm r}_{\iota}$ lies in a reference magnetic cell and
$\bm L_{AB}=A\tilde{\bm a}_1+B\tilde{\bm a}_2$, with $A,B\in\mathbb Z$. The coordinate sewing relations give
\begin{equation}
    \mathcal U^S_{\bm k}(r;\tilde r_{\iota})
    =e^{i\bm k\cdot\bm L_{AB}}
    e^{i[\bm L_{AB}\times\tilde{\bm r}_{\iota}
    -AB\tilde{\bm a}_1\times\tilde{\bm a}_2]/(2\ell_B^2)}.
    \label{eq:rf_extension}
\end{equation}
The cocycle is $e^{-i\pi pAB}$, reducing to $e^{-i\pi AB}$ at $p=1$.

On a finite torus containing $\mathcal N_x\mathcal N_y=N_s/q$
magnetic cells, the unitary magnetic Fourier transform has normalization
$(\mathcal N_x\mathcal N_y)^{-1/2}$.
For $\bm r=\tilde{\bm r}_{\iota}+\bm L_{AB}$ and
$\bm r'=\tilde{\bm r}_{\iota'}+\bm L_{A'B'}$, the hopping is
\begin{equation}
    J_{\bm r,\bm r'-\bm r}
    =\frac1{\mathcal N_x\mathcal N_y}\sum_{\bm k\in{\rm MBZ}_T}
    \mathcal U^S_{\bm k}(r';\tilde r_{\iota'})
    (h_{\bm k})_{\iota'\iota}
    \mathcal U^S_{\bm k}(r;\tilde r_{\iota})^*.
    \label{eq:rf_full_hopping}
\end{equation}
Here $\mathcal U^S_{\bm k}$ is defined in Eq.~\eqref{eq:rf_extension}.
In the infinite-lattice limit, the normalized momentum sum becomes
the normalized MBZ integral.

For torus periods $\bm L_1=\mathcal N_x\tilde{\bm a}_1$
and $\bm L_2=\mathcal N_y\tilde{\bm a}_2$, the boundary twists of
Sec.~\ref{app:KM_gauge} select the allowed momenta through
$\bm k\cdot\bm L_\mu=\vartheta_\mu\pmod{2\pi}$, $\mu=1,2$.

In a compatible periodic magnetic-cell basis, analytic continuation of $h_{\bm k}$ to a strip of nonzero width in each complex momentum component implies
exponentially decaying real-space hopping
\cite{ChenMazaheriSeidelTang2014,Read2017CompactWannier}.
\endgroup

For the state $\Psi_{mn}$ in
Eq.~\eqref{eq:parton_chim_chin}, take both Slater determinants at particle number $N$, with their effective fields summing to the physical field. Their torus fluxes are the integers $N/m$ and $N/n$, with $N_\phi=\phi N_s=N/m+N/n$, and their boundary phases are chosen so that their product satisfies the required torus boundary conditions. The magnetic lengths $\ell_m$ and $\ell_n$ associated with $\chi_m$ and $\chi_n$ satisfy $\ell_m^{-2}+\ell_n^{-2}=\ell_B^{-2}$, so their Gaussian factors combine into $e^{-\sum_i|z_i|^2/(4\ell_B^2)}$. After removing the Gaussian factors,
Eq.~\eqref{eq:rf_ll_ladder} bounds the degree in each $\bar z_i$
by $m-1$ and $n-1$ in the two determinants. Their product therefore
lies in the first $m+n-1$ LLs. Choosing $n_{\rm LL}=m+n-1$ in
Eq.~\eqref{eq:rf_positive_block} gives the sampled state zero kinetic energy,
while its coincidence zeros give zero onsite interaction energy~\cite{CruiseSeidel2026,Ahari2023Fibonacci}.

{

\Needspace{18\baselineskip}
\paragraph{\texorpdfstring{{Single higher-Chern bands and Halperin-type states}}{Single higher-Chern bands and Halperin-type states}.}
\label{sec:HC_single_band}
\begingroup

We use the non-Bravais geometry, bond labels, and site modulation of
Sec.~\ref{app:nonbravais_spin}, with $n_{\rm sub}\ge3$ sites per primitive cell.
Here $\bm k$ labels the ordinary Bloch phase $e^{i\bm k\cdot\bm R}$
of the modulated sampled-LL states, and $\bm a_1\times\bm a_2=2\pi\ell^2$.
Following Li and Wu
\cite{LiWu2026LatticeGLL}, let
\begin{equation}
 V_{\bm k}^{\rm NB}
 =\bigl(|e_{0,\bm k}\rangle,\ldots,
 |e_{n_{\rm sub}-2,\bm k}\rangle\bigr)
 \label{eq:HC_NB_frame}
\end{equation}
collect the cell amplitudes of the first $n_{\rm sub}-1$ modulated and
sampled LL states in the cell Bloch convention of
Sec.~\ref{app:nonbravais_spin}, with components
$[e_{n,\bm k}]_\alpha=\mathcal B_\alpha\psi_{n,\bm k}(\bm\tau_\alpha)$,
where $\alpha=1,\ldots,n_{\rm sub}$ and $n=0,\ldots,n_{\rm sub}-2$.
Each column belongs to $\mathbb C^{n_{\rm sub}}$.
Assume that $V_{\bm k}^{\rm NB}$ has full column rank throughout the Brillouin zone.

The orthogonal projector onto the sampled-LL span is
\begin{equation}
 Q_{\bm k}^{\rm LL}=V_{\bm k}^{\rm NB}
 \bigl[(V_{\bm k}^{\rm NB})^\dagger V_{\bm k}^{\rm NB}\bigr]^{-1}
 (V_{\bm k}^{\rm NB})^\dagger.
 \label{eq:HC_LL_projector}
\end{equation}
The sampled-LL subspace has total Chern number $n_{\rm sub}-1$, so its one-dimensional complement has $C=-(n_{\rm sub}-1)$.

Li and Wu show that the LL derivative structure makes the
complementary-band Bloch state anti-holomorphic in momentum up to a
common scalar factor, yielding $\operatorname{tr}g_{\bm k}=|\Omega_{\bm k}|$
for its quantum metric $g_{\bm k}$ and Berry curvature $\Omega_{\bm k}$ \cite[Supplemental Material III, Eqs.~(S12)--(S19)]{LiWu2026LatticeGLL}. This statement uses Bloch phases referred to the physical orbital positions, related to our cell Fourier convention below.

We choose the positive-semidefinite Hamiltonian
\begin{equation}
    h_{\bm k}^{(C)}=E_{\bm k}Q_{\bm k}^{\rm LL},\qquad E_{\bm k}>0.
    \label{eq:HC_parent}
\end{equation}
It acts in the $n_{\rm sub}$-dimensional physical sublattice space and places the higher-Chern band at zero energy, while the $n_{\rm sub}-1$ LL-like bands share the positive dispersion $E_{\bm k}$.
Choosing $E_{\bm k}$ to be constant also flattens the positive-energy bands. More generally, let $|\Phi_{n,\bm k}\rangle$ denote the orthonormal states obtained by Gram--Schmidt orthogonalization of the
columns of $V_{\bm k}^{\rm NB}$. Independent positive dispersions give
\begin{equation}
    h_{\bm k}^{(C)}=\sum_{n=0}^{n_{\rm sub}-2}
    \varepsilon_n(\bm k)|\Phi_{n,\bm k}\rangle\langle\Phi_{n,\bm k}|,
    \qquad\varepsilon_n(\bm k)>0.
    \label{eq:HC_positive_dispersions}
\end{equation}
These choices have the same zero-energy band and ideal geometry, but generally have different hopping amplitudes and positive-energy spectra. For exponentially decaying hopping, the energy functions are chosen so that $h_{\bm k}^{(C)}$ is periodic and admits analytic continuation to a complex neighborhood of the real Brillouin zone, in the cell Fourier convention below.

On a torus of $N_{\rm cell}$ primitive cells, choose the cell Fourier
convention $\langle\bm R+\bm\tau_\alpha|\bm k,\beta\rangle
=N_{\rm cell}^{-1/2}e^{i\bm k\cdot\bm R}\delta_{\alpha\beta}$.
Li and Wu instead use orbital-embedded Bloch phases
$e^{i\bm k\cdot(\bm R+\bm\tau_\alpha)}$
\cite[Supplemental Material, Eq.~(S10)]{LiWu2026LatticeGLL}.
For a band eigenvector, the coefficients in the two conventions obey
$u_{\bm k,\alpha}^{\rm emb} = e^{-i\bm k\cdot\bm\tau_\alpha}u_{\bm k,\alpha}^{\rm cell}$. The ideal-geometry statement above uses the embedded coefficients; the following hopping transform uses cell coefficients. The real-space couplings are
\begin{equation}
    W_{\alpha\beta}^{(C)}(\bm D)
    e^{i\varphi_{\alpha\beta}^{(C)}(\bm D)}
    =\frac1{N_{\rm cell}}\sum_{\bm k}e^{i\bm k\cdot\bm D}
    [h_{\bm k}^{(C)}]_{\alpha\beta},
    \label{eq:HC_sublattice_couplings}
\end{equation}
where the momentum sum obeys the prescribed boundary twists, and
$W_{\alpha\beta}^{(C)}(\bm D)\ge0$. The corresponding spin Hamiltonian is
\begin{equation}
    H_{\rm XY}^{(C)}
    =\sum_{\bm R,\bm D}\sum_{\alpha,\beta=1}^{n_{\rm sub}}
    W_{\alpha\beta}^{(C)}(\bm D)
    e^{i\varphi_{\alpha\beta}^{(C)}(\bm D)}
    S_{\bm R+\bm D+\bm\tau_\alpha}^{+}
    S_{\bm R+\bm\tau_\beta}^{-}.
    \label{eq:HC_sublattice_spin_parent}
\end{equation}
This is the sublattice-resolved form of
Eq.~\eqref{eq:Hspin_general_nLL}. The directed sum includes the onsite terms at $\bm D=\bm0$, $\alpha=\beta$. Their coefficients are
$\epsilon_{\bm R+\bm\tau_\alpha}=\epsilon_\alpha
=N_{\rm cell}^{-1}\sum_{\bm k}[h_{\bm k}^{(C)}]_{\alpha\alpha}$,
which may differ between sublattices.

Color-entangled constructions relate ideal higher-Chern bands to multicomponent Landau levels~\cite{WuRegnaultBernevig2013,WangKlevtsovLiu2023Ideal,DongLedwith2023Ideal}. In lattice models with a higher-Chern flat band and two-body on-site repulsion, bosonic Halperin-type states can be exact zero-energy ground states at band filling $1/(|C|+1)$~\cite{BehrmannLiuBergholtz2016,WangKlevtsovLiu2023Ideal}.

Our rank-one band has one orbital per primitive cell and the lattice has $n_{\rm sub}$ spin-$1/2$ sites per cell. At the Halperin filling, we have
\begin{equation}
 \nu_b=\frac{N}{N_{\rm cell}}=\frac1{|C|+1},
 \qquad m_0=\frac{2\nu_b}{n_{\rm sub}}-1,
 \label{eq:HC_filling_magnetization}
\end{equation}
where $N_{\rm cell}$ is the number of primitive cells and $N_s=n_{\rm sub}N_{\rm cell}$ is the number of spins. For $n_{\rm sub}=3$, the target band has $C=-2$, and $\nu_b=1/3$ corresponds to $m_0=-7/9$. On the 54-site torus, we find three zero modes. The particle entanglement spectrum of their equal-weight mixture is consistent with Halperin $(221)$ topological order (Appendix~\ref{app:numerics}, Fig.~\ref{fig:highchern54-spectrum-pes}).
\endgroup

\par}

\subsection{Density interactions between distinct sites and exact states}
\label{app:positive_delta_interactions}
{

For Route 2, we express $H_{\rm XY}+H_{2,\delta}^{(n_2)}$ and its
higher-body extension in Eq.~\eqref{eq:spin_half_deformed_parent} as sums of positive-semidefinite terms, then construct deformed Laughlin and
dressed Moore--Read zero modes on the infinite lattice.

\paragraph{\texorpdfstring{Displaced density interactions and zero-mode conditions}
{Displaced density interactions and zero-mode conditions}.}
The local degrees of freedom are spin \(1/2\), with \(n_{\bm r}=S^z_{\bm r}+1/2\), so the onsite hard-core constraint is built into the Hilbert space. The \(n_2\) off-site pair displacements are nonzero and distinct up to inversion; all interaction coefficients are positive and each cluster contains distinct lattice sites. Define,
\begin{equation}
B_{\bm r,\lambda}^{(2)}
:=S_{\bm r}^{-}S_{\bm r+\bm\delta_\lambda}^{-},
\qquad
B_{\bm r,\delta'}^{(k+1)}
:=\prod_{\ell=0}^{k}S_{\bm r+\ell\bm\delta'}^{-}.
\label{eq:app_deformed_cluster_lowering}
\end{equation}
For distinct sites, \(B^\dagger B\) is exactly the corresponding
product of densities. In the occupation basis, \(B|\Psi\rangle=0\)
means that the amplitude vanishes whenever all sites in that cluster
are occupied.

For a finite geometry, let \(w_\gamma\) be orthonormal positive-energy
eigenmodes of the one-body hopping, with \(\epsilon_\gamma>0\), and set
$A_\gamma:=\sum_{\bm r}w_\gamma(\bm r)^*S_{\bm r}^{-}$.
In the zero-energy normalization of Eq.~{\eqref{eq:H_XY_intro}},
including its onsite diagonal shift,
\(H_{\rm XY}=\sum_\gamma\epsilon_\gamma A_\gamma^\dagger A_\gamma\).
Consequently,

\begin{equation}
H_{\rm def}^{(k,n_2)}
=\sum_\gamma\epsilon_\gamma A_\gamma^\dagger A_\gamma
+\sum_{\bm r}\sum_{\lambda=1}^{n_2}
\mathcal V_{\bm\delta_\lambda}^{(2)}
\left(B_{\bm r,\lambda}^{(2)}\right)^\dagger
B_{\bm r,\lambda}^{(2)}
+\mathcal V_{\bm\delta'}^{(k+1)}\sum_{\bm r}
\left(B_{\bm r,\delta'}^{(k+1)}\right)^\dagger
B_{\bm r,\delta'}^{(k+1)}
\succeq0 .
\label{eq:app_deformed_full_positive_parent}
\end{equation}
On the infinite lattice the mode sum is understood as the corresponding
spectral integral.
 {Any nonzero state annihilated by all $A_\gamma$ and all relevant cluster
operators $B$ is therefore an exact zero-energy ground state.} For the pair-interaction Hamiltonian,
the last term in Eq.~\eqref{eq:app_deformed_full_positive_parent} is omitted. For $k=1$, the cluster term is itself a pair interaction and is included among the $n_2$ distinct pair displacements. Repeated displacements
combine their couplings without increasing $n_2$.

On a finite geometry, an optional uniform \(H_Z=-hS^z_{\rm tot}\)
shifts sector energies but is not part of the positive-semidefinite
decomposition above.

\paragraph{Deformed Laughlin wavefunctions.}
For the exact KM kernel with $0<\phi<1$ and fixed
particle number $N$, let $z_{ij}=z_i-z_j$, and let
$\delta_\lambda$ be the complex coordinate of the nonzero lattice
displacement $\bm\delta_\lambda$.
For displacements distinct up to inversion, an explicit amplitude is
\cite{JiWang2026}
\begin{equation}
    \Phi_{\rm L}^{\{\delta\}}(\{z\}) \propto\prod_{i<j}\left[z_{ij}^{2} \prod_{\lambda=1}^{n_2}(z_{ij}^{2}-\delta_\lambda^{2})\right] e^{-\frac{\phi}{4\ell^2}\sum_i|z_i|^2}. \label{eq:deformed_Laughlin_plane}
\end{equation}

The additional pair factors dress the Laughlin amplitude with
zeros at the prescribed separations $\pm\delta_\lambda$, preserving bosonic symmetry and the quadratic coincidence zeros. Multiplication by $\prod_i\eta(z_i)$, as in Eq.~\eqref{eq:Psi_KL_spin}, places each coordinate in the exact kinetic kernel. At fixed $N$, its nonzero polynomial factor remains nonzero upon infinite-lattice sampling and the Gaussian makes the norm finite. The corresponding spin state is therefore a zero mode of $H_{\rm XY}+H_{2,\delta}^{(n_2)}$. The pair degree $2(n_2+1)$ identifies the deformed Laughlin sequence $\nu_b=1/[2(n_2+1)]$; $n_2=0$ recovers the original Laughlin amplitude. Here $\Phi_{\rm L}^{\{\delta\}}$ includes the Gaussian, as in Eq.~\eqref{eq:PhiL_continuum}; the MR polynomial below does not.

With the interaction projected to the KM flat band, finite-size scaling of the many-body gap at $\nu_b=1/4$ and $\phi=1/3$ supports a nonzero gap in the thermodynamic limit~\cite{JiWang2026}.

\paragraph{\texorpdfstring{Dressed Moore--Read wavefunction} {Dressed Moore--Read wavefunction}.}

We give the lattice amplitudes of the dressed Moore--Read states introduced in the main text and prove their zero-energy property.

For the KM Hamiltonian with $0<\phi<1$ and fixed
even $N$, use $z_{ij}$ and $\delta_\lambda$ as above, and let
$\delta'$ be the complex coordinate of $\bm\delta'$.
Define $\mathcal J:=\prod_{i<j}z_{ij}$ and
$A_{\delta'}(z):=z^2-(\delta')^2$.
The pair factor $z_{ij}^2\prod_{\lambda=1}^{n_2}(z_{ij}^2-\delta_\lambda^2)$
has degree $2+2n_2$, with a quadratic coincidence zero and simple zeros
at $z_{ij}=\pm\delta_\lambda$. Define
\begin{equation}
    \mathcal P_{\rm dMR}^{\delta',\{\delta\}}(\{z\})
    :=\operatorname{Pf}\!\left[
    \frac{z_{ij}}{A_{\delta'}(z_{ij})}\right]
    \,\prod_{i<j}
    \left[z_{ij}A_{\delta'}(z_{ij})
    \prod_{\lambda=1}^{n_2}(z_{ij}^2-\delta_\lambda^2)\right].
    \label{eq:app_dressed_MR_plane}
\end{equation}
For $n_2=0$, this reduces to the polynomial
in Eq.~\eqref{eq:dressed_MR_main}.
The Pfaffian is of the antisymmetric \(N\times N\) matrix with the
displayed off-diagonal entries and zero diagonal. At apparent poles,
Eq.~\eqref{eq:app_dressed_MR_plane} means the polynomial obtained
by the denominator-cleared expansion below.

The corresponding lattice amplitude, including the Gaussian and
the site gauge of Eq.~\eqref{eq:Psi_KL_spin}, is
\begin{equation}
\Psi_{\rm dMR}(\bm r_1,\ldots,\bm r_N)
={\frac{1}{\sqrt{\mathcal N_{\rm dMR}}}}
\left[\prod_{i=1}^{N}\eta(z_i)\right]
e^{-\frac{\phi}{4\ell^2}\sum_{i=1}^{N}|z_i|^2}
\mathcal P_{\rm dMR}^{\delta',\{\delta\}}(\{z\}),
\qquad z_i=z(\bm r_i).
\label{eq:app_dmr_infinite_lattice_amplitude}
\end{equation}

Here \(1/\sqrt{\mathcal N_{\rm dMR}}\) is the normalization factor.
The associated ket has coefficient \(\Psi_{\rm dMR}(\bm r_1,\ldots,\bm r_N)\) on \(\prod_iS^+_{\bm r_i}|\Downarrow\rangle\).

Let \(\mathcal M_N\) be the perfect matchings of the particle labels:
each matching partitions \(\{1,\ldots,N\}\) into \(N/2\) disjoint
unordered pairs, with each label appearing exactly once.
Writing each pair as \(i<j\), let \(\epsilon(\pi)\) be its Pfaffian
permutation sign. Equation~\eqref{eq:app_dressed_MR_plane} expands as
\begin{equation}
    \mathcal P_{\rm dMR}^{\delta',\{\delta\}} = \mathcal J
    \prod_{i<j}\prod_{\lambda=1}^{n_2} (z_{ij}^2-\delta_\lambda^2)
    \,\sum_{\pi\in\mathcal M_N}\epsilon(\pi) \prod_{(i,j)\in\pi}z_{ij}
    \prod_{\substack{i<j\\(i,j)\notin\pi}}A_{\delta'}(z_{ij}). \label{eq:app_dmr_matching_expansion}
\end{equation}

In each matching term, a paired factor
\(A_{\delta'}(z_{ij})\) is canceled by its Pfaffian denominator, leaving \(z_{ij}\); every unpaired factor \(A_{\delta'}(z_{ij})\) remains. The matching sum is an alternating polynomial: exchanging two particle labels changes the Pfaffian sign while leaving
\(\prod_{i<j}A_{\delta'}(z_{ij})\) invariant. It is therefore divisible by \(\mathcal J\), so the full amplitude contains \(\mathcal J^2\) and is symmetric, with at least quadratic coincidence zeros. The outer factors enforce all prescribed off-site pair zeros.

Choose arbitrary distinct particle labels \(a,b,c\),
and set \(z_a=z\), \(z_b=z+\delta'\), and \(z_c=z+2\delta'\), with
\(\delta'\ne0\), while leaving all other coordinates arbitrary.
Then \(A_{\delta'}(z_{ab})=A_{\delta'}(z_{bc})=0\), whereas
\(z_{ab}=z_{bc}=-\delta'\ne0\). A perfect matching cannot pair \(b\)
with both \(a\) and \(c\); hence at least one of these two vanishing
\(A_{\delta'}\) factors remains in every term. The full amplitude
therefore satisfies the displaced three-particle constraint.

The highest-degree part of Eq.~\eqref{eq:app_dmr_matching_expansion} is the nonzero polynomial $\operatorname{Pf}[1/z_{ij}]\,\mathcal J^{2n_2+3}$, equal to $\mathcal J^{2n_2+2}$ times the bosonic MR polynomial~\cite{MooreRead1991,ReadRezayi1999,BoruttaFeuerpfeilIqbal2026}. Its degree in each particle coordinate is $(2n_2+3)(N-1)-1$, giving the filling $\nu_b=1/(2n_2+3)$ in the thermodynamic limit.
Sampling on the full infinite lattice cannot make a nonzero polynomial identically zero, and the Gaussian makes its norm finite at fixed \(N\). With all other coordinates fixed, each particle coordinate has the gauge-transformed sampled-LLL form of the exact kinetic kernel. The kinetic term therefore annihilates the state. Together with the pair and three-particle zeros proved above, the positive decomposition in Eq.~\eqref{eq:app_deformed_full_positive_parent} establishes an exact zero-energy ground state on the infinite lattice. On a finite geometry, the amplitude must additionally belong to the
actual one-body kernel with its prescribed boundary conditions.

For \(n_2=0\), Fig.~\ref{fig:rr_supp}(e,f) in Appendix~\ref{app:numerics} shows
finite-torus spectra and PES at \(\nu_b=N/(\phi N_s)=1/3\),
with a nine-state zero-mode manifold and counting consistent with
the dressed MR state.
}

\subsection{Higher-spin Hamiltonians and Read--Rezayi states}
\label{app:spin_truncated_boson}
\begingroup
\setlength{\abovedisplayskip}{8pt plus 2pt minus 2pt}
\setlength{\belowdisplayskip}{8pt plus 2pt minus 2pt}

Keeping the KM hopping, we obtain higher-spin Read--Rezayi parent Hamiltonians by replacing the onsite two-body repulsion with a $(k+1)$-body repulsion and taking its hard-core limit. Related spin-$1/2$ cluster constructions realize the $k=2,3$ Read--Rezayi Hamiltonians in a strong-coupling limit~\cite{KapitSimon2013}. Here each constrained boson site is represented by a single spin $S=k/2$.

For an integer $k\ge1$, we take the bosonic Hamiltonian
\[
 H_{\rm KM}^{(k+1)}=H_{\rm KM}
 +U_{k+1}\sum_{\bm r}\binom{\hat n_{\bm r}}{k+1},
 \qquad \hat n_{\bm r}=c_{\bm r}^{\dagger}c_{\bm r},\quad U_{k+1}>0 .
\]
For $k=1$ and $U_2=U$, this gives the Hamiltonian in Eq.~\eqref{eq:HBH_def}. The lattice-sampled $Z_k$ Read--Rezayi state is an exact zero-energy ground state at $\nu_b=k/2$, provided that lattice sampling preserves the continuum zero-mode structure~\cite{MooreRead1991,ReadRezayi1999,BehrmannLiuBergholtz2016}.

In the strict limit $U_{k+1}\to\infty$, the local boson occupation is restricted to $n=0,\ldots,k$, giving the Hilbert space $\mathcal H_k=\operatorname{span}\{|n\rangle_b:n=0,1,\ldots,k\}$. We identify these occupation states with the $S^z$ eigenstates of a spin $S=k/2$:
\begin{equation}
  |n\rangle_b \longleftrightarrow |m=n-S\rangle_s,
  \qquad \hat n=S^z+S .
  \label{eq:app_spin_boson_hilbert}
\end{equation}
The constrained operators $a^\dagger$ and $a$ retain the bosonic matrix elements within $\mathcal H_k$ and satisfy $a^\dagger|k\rangle=a|0\rangle=0$. Define $(k-\hat n)_{\rm mp}^{-1/2}$ to act as $(k-n)^{-1/2}$ on $|n\rangle$ for $n<k$ and as zero on $|k\rangle$; the subscript $\mathrm{mp}$ denotes the Moore--Penrose inverse~\cite{Penrose1955}. The inverse Holstein--Primakoff representation restricted to $\mathcal H_k$~\cite{HolsteinPrimakoff1940} is
\begin{equation}
  a^\dagger=S^+(k-\hat n)_{\rm mp}^{-1/2},
  \qquad
  a=(k-\hat n)_{\rm mp}^{-1/2}S^- .
  \label{eq:app_canonical_boson_spin}
\end{equation}
For $k=1$, Eq.~\eqref{eq:app_canonical_boson_spin} reduces to the Matsubara--Matsuda map $a^\dagger=S^+$ and $a=S^-$. Writing $F_{\bm r}^{(k+1)}:=(S-S_{\bm r}^z)_{\rm mp}^{-1/2}$, the hopping becomes
\begin{equation}
  a_{\bm r+\bm d}^\dagger a_{\bm r}
  =
  S_{\bm r+\bm d}^{+}F_{\bm r+\bm d}^{(k+1)}
  F_{\bm r}^{(k+1)}S_{\bm r}^{-}.
  \label{eq:app_canonical_hopping_spin}
\end{equation}
The onsite interaction vanishes on the constrained subspace, and the restriction of $H_{\rm KM}^{(k+1)}$ gives
\begin{equation}
  H_{\rm RR}^{(S)}
  =
  \sum_{\bm r,\bm d\in\Lambda}
  W_{\bm d}e^{i\varphi_{\bm r,\bm d}}
  S_{\bm r+\bm d}^{+}F_{\bm r+\bm d}^{(k+1)}
  F_{\bm r}^{(k+1)}S_{\bm r}^{-}.
  \label{eq:canonical_spinS_parent}
\end{equation}
The mapped Read--Rezayi state has zero energy and is therefore an exact ground state of the positive-semidefinite spin Hamiltonian in the corresponding magnetization sector. With $M=S^z_{\rm tot}=N-SN_s$, its normalized magnetization at $\nu_b=k/2$ is $m_0^{(S)}:=M/(SN_s)=\phi-1$. The cases $k=1,2,3$ give the Laughlin/KL spin-$1/2$, Moore--Read spin-1, and $Z_3$ Read--Rezayi spin-$3/2$ parent Hamiltonians, respectively.

\endgroup

\begingroup
\section{Finite-size energy spectra and particle entanglement spectra}
\label{app:numerics}
\label{app:parton_numerics}
\label{app:rr_numerics}

For representative spin models constructed in Appendix~\ref{app:generalizations}, we report finite-size excitation gaps and compare the momentum-resolved level counts in the particle entanglement spectrum (PES) with theoretical predictions. For the Read--Rezayi and dressed Moore--Read models, we compare the PES counts with those obtained from the corresponding generalized Pauli principles (GPPs). For the models with parton ground states $\chi_m\chi_n$, we derive a counting rule from the entangled Pauli principle (EPP) and also obtain independent PES counts directly from the parton wavefunctions. We test the EPP prediction for the $\chi_3^2$ model against its finite-torus quasihole zero-mode count.

\subsection{Numerical setup and observables}
\label{app:numerical_setup}
We consider the parton models of Sec.~\ref{app:projector_mll}, the higher-spin models of Sec.~\ref{app:spin_truncated_boson}, and the dressed Moore--Read model of Sec.~\ref{app:positive_delta_interactions} on triangular tori with zero boundary twists. The torus contains $N_s=L_1L_2$ sites, with $L_1,L_2$ specifying its size along the primitive vectors defined in Sec.~\ref{app:KL_review}. The corresponding magnetic momentum grid contains $\mathcal N_x\times\mathcal N_y$ points, and $N_\phi=\phi N_s$ is the number of flux quanta. 
Table~\ref{tab:numerical_parameters} lists the parameters used for these calculations.

To obtain the PES, we trace out $N-N_A$ particles from the equal-weight mixture of the $g$ orthonormal ground states~\cite{LiHaldane2008,
SterdyniakRegnaultBernevig2011,SterdyniakRegnaultMoller2012}:
\begin{equation}
 \rho_A=\operatorname{Tr}_{N-N_A}\!\left(
 \frac1g\sum_{a=1}^{g}|\Psi_a\rangle\langle\Psi_a|
 \right).
 \label{eq:supp_pes_definition}
\end{equation}
The entanglement energies are $\xi=-\ln\lambda$, where $\lambda$
is an eigenvalue of $\rho_A$.
Subsystem translation sectors are labeled by
$K_A=K_y^A+\mathcal N_yK_x^A$, with momentum components $K_x^A,K_y^A$.

Let $H_{N_A}$ be the same Hamiltonian restricted to $N_A$ particles,
at fixed geometry, flux, and boundary conditions.
For the exact zero modes considered here, particle removal preserves
the defining annihilation constraints. The PES support therefore
satisfies $\operatorname{supp}\rho_A\subseteq\ker H_{N_A}$, and hence
$\operatorname{rank}\rho_A\leq\dim\ker H_{N_A}$.
Equality requires the particle contractions of the chosen
ground-state manifold to span this zero-mode space.

Let $T_x,T_y$ translate all particles by the two magnetic-cell vectors. Consistently with Sec.~\ref{app:projector_mll}, we label their simultaneous eigenstates by,
\begin{equation}
     T_\mu|\Psi_{\bm K}\rangle
     =e^{-2\pi iK_\mu/\mathcal N_\mu}|\Psi_{\bm K}\rangle,\qquad
     K_\mu=0,\ldots,\mathcal N_\mu-1,\quad \mu=x,y.
     \label{eq:MP_translation_sectors}
\end{equation}
The subsystem translations $T_\mu^A$ use the same convention for $K_\mu^A$. Blue PES points have $\lambda>\tau_{\rm PES}=10^{-14}$. Gray circles and crosses mark small positive and negative eigenvalues, respectively, that are treated as numerical zeros and plotted at $-\ln|\lambda|$. Dashed lines mark $-\ln\tau_{\rm PES}$. An exact zero eigenvalue corresponds to $\xi=\infty$.

In Figs.~\ref{fig:parton_supp}--\ref{fig:rr_supp}, we plot the energies without subtracting the ground-state energy. For each energy $E$ shown in this appendix, the corresponding normalized state $|\psi\rangle$ satisfies $\lVert H|\psi\rangle-E|\psi\rangle\rVert\leq10^{-8}$ in the energy units used for diagonalization.

\begin{center}
\begin{minipage}{\linewidth}
\captionof{table}{
Parameters for Figs.~\ref{fig:parton_supp}--\ref{fig:rr_supp}. $H_{mn}$ denotes the parent Hamiltonian for the parton state $\chi_m\chi_n$.
The physical torus has $N_s=L_1L_2$ sites; $n_{\rm LL}$ counts the Landau levels sampled on the lattice, and $g$ is the ground-manifold dimension. For spin $S$ on each site, $N$ fixes the magnetization sector through $S_{\rm tot}^z=N-SN_s$.}
\label{tab:numerical_parameters}
\label{tab:parton_parameters}
\label{tab:rational_parameters}
\centering\small
\setlength{\tabcolsep}{5pt}
\renewcommand{\arraystretch}{1.14}
\begin{tabular}{lcccccccc}
\hline
Model & $L_1\!\times\!L_2$ & $\phi$ & $N$ & $N_\phi$ & $n_{\rm LL}$ &
$(\mathcal N_x,\mathcal N_y)$ & $N_A$ & $g$\\
\hline
$H_{21}$ & $6\times6$ & $1/4$ & 6 & 9 & 2 & $(3,3)$ & 3 & 3\\
$H_{22}$ & $6\times6$ & $1/6$ & 6 & 6 & 3 & $(2,3)$ & 3 & 3\\
$H_{33}$ & $6\times8$ & $1/12$ & 6 & 4 & 5 & $(1,4)$ & 4 & 4\\
\hline
$H_{21}$ & $5\times6$ & $3/10$ & 6 & 9 & 2 & $(1,3)$ & 3 & 3\\
$H_{22}$ & $8\times4$ & $3/16$ & 6 & 6 & 3 & $(2,1)$ & 3 & 3\\
\hline
Spin-1 MR & $4\times4$ & $1/2$ & 8 & 8 & 1 & $(2,4)$ & 4 & 3\\
Spin-$3/2$ $Z_3$ RR & $4\times3$ & $1/2$ & 9 & 6 & 1 & $(2,3)$ & 4 & 4\\
Dressed MR & $8\times4$ & $3/4$ & 8 & 24 & 1 & $(2,4)$ & 4 & 9\\
\hline
\end{tabular}
\end{minipage}
\end{center}

For the spin-$1/2$ models $H_{mn}$, we obtain the exchange couplings from $h_{\bm k}=(\det G_{\bm k})P^\perp_{\bm k}$ as in Sec.~\ref{app:projector_mll} and retain exchanges at all distances. The target parton states are $\chi_2\chi_1$, $\chi_2^2$, and $\chi_3^2$. Additional many-body zero modes beyond the target parton manifold can occur on particular finite clusters~\cite{CruiseSeidel2026}. For the $p=1$ examples in Fig.~\ref{fig:parton_supp}, we displace all sampling points by $\bm\rho_0=(\pi a_0/4)\hat{\bm y}$, where $a_0=|\bm a_1|$ and $\bm a_1\parallel\hat{\bm x}$, using the sampling convention of Ref.~\cite{CruiseSeidel2026} adapted to the triangular lattice.
Zero boundary twists are specified relative to this sampling convention
and the magnetic-translation convention of Sec.~\ref{app:projector_mll}. The $p>1$ examples use magnetic cells
$(\kappa_1,\kappa_2)=(5,2),(4,4)$, respectively.

For the spin-1 Moore--Read and spin-$3/2$ $Z_3$ Read--Rezayi models, we retain the full KM hopping and impose the onsite constraints $n_r\le2$ and $n_r\le3$, respectively, as in Sec.~\ref{app:spin_truncated_boson}.

For the dressed Moore--Read model, we retain the KM exchanges and exclude configurations with three consecutive up spins along $\bm a_1$. This corresponds to $n_2=0$ and $\bm\delta'=\bm a_1$ in Sec.~\ref{app:positive_delta_interactions}, giving $\nu_b=1/3$.

\subsection{PES and zero-mode counting}
\label{app:pes_counting}
We use momentum-resolved PES and zero-mode counting to test the candidate CSL topological orders in the spin models. For the torus counts, let $j=0,\ldots,L-1$ label the guiding-center orbitals, with $L=N_\phi$. Their occupations satisfy $n_j\in\mathbb Z_{\ge0}$ and $\sum_j n_j=N_A$, with indices taken modulo $L$. These $n_j$ are orbital
occupations; the site occupations used in the lattice Hamiltonians
are denoted by $n_r$. For the spin-1 MR and spin-$3/2$ $Z_3$ RR models, the GPP count
assigns unit weight to each pattern satisfying
$n_j+n_{j+1}\le k$, with $k=2$ for MR and $k=3$ for $Z_3$
\cite{MooreRead1991,ReadRezayi1999,BernevigHaldane2008Jack}.
For dressed MR ($L=24$), the two restrictions are
\begin{equation}
 n_j+n_{j+1}\le1,\qquad
 \sum_{a=0}^{5}n_{j+a}\le2.
 \label{eq:supp_dressed_gpp}
\end{equation}

For the parton models, each root occupation pattern contributes once for each independent internal LL state allowed by the EPP~\cite{Bandyopadhyay2018EPP,Bandyopadhyay2020Jain}.
We first derive the local constraints on a cylinder, where orbitals have a linear ordering and each orbital has $n_{\rm LL}$ internal states. At fixed total orbital index $J=\sum_j j n_j$, a root is an occupation pattern maximal under dominance among the patterns with nonzero coefficients in the state. Dominance is generated by inward squeezing.

We represent each particle's $n_{\rm LL}$ internal coefficients by
homogeneous polynomials of degree $r:=n_{\rm LL}-1$ in auxiliary variables $\bm s_a=(u_a,v_a)^{\mathsf T}$, where $a$ labels particles, and write $[ab]=u_av_b-v_au_b$ for the antisymmetric pair factor
\cite{Ahari2023Fibonacci}. The single-particle polynomial basis is
$\{u_a^{r-\alpha}v_a^\alpha\}_{\alpha=0}^{r}$.
To obtain the local factors, consider two-particle roots of the continuum contact problem. In a polynomial LL basis, let
$Q_{jk}(\bm s_a,\bm s_b)$ denote the internal coefficient polynomial
for guiding-center orbitals $j\leq k$. It has degree $r$ in each
particle's spinor separately. At coincidence, pairs with the same total
orbital index $j+k$ share a common orbital factor, so their diagonal
polynomials contribute to the same contact constraint.
We absorb the orbital normalization and bosonic symmetrization factors
into $Q_{jk}$, so that the zero-mode condition at fixed $J$ reads
\begin{equation}
    \sum_{\substack{j\leq k\\j+k=J}}
    Q_{jk}(\bm s,\bm s)=0
    \qquad\text{for every }\bm s.
    \label{eq:supp_weighted_contact_constraint}
\end{equation}
The inward-squeezed pairs available to a pair $(j,k)$ are
\begin{equation}
    (j,k)\longmapsto(j+s,k-s),\qquad
    s\in\mathbb Z_{>0},\quad j+s\leq k-s.
    \label{eq:supp_weighted_inward_squeezing}
\end{equation}
Same-orbital and neighboring-orbital pairs admit no such squeezing.
If such a pair is a two-particle root, all more outward pairs with the
same total orbital index are absent. With no inward pair available
either, its coefficient must satisfy $Q_{jk}(\bm s,\bm s)=0$ for every
auxiliary spinor $\bm s$, which requires a factor $[ab]$.
For two bosons in the same orbital, $Q_{jj}$ is symmetric under particle
exchange; dividing by $[ab]$ leaves an antisymmetric polynomial and
hence requires a second factor:
\begin{equation}
    Q_{j,j}=[ab]^2R_{j,j},\qquad
    Q_{j,j+1}=[ab]R_{j,j+1},
    \label{eq:supp_weighted_bracket_factors}
\end{equation}
where $R$ denotes the remaining polynomial.

For $k-j\geq2$, an inward-squeezed pair with the same total orbital
index can cancel the root's coincidence amplitude. For any $Q_{jk}$,
choose the center pair and its polynomial as
\begin{equation}
    \begin{aligned}
    (j_*,k_*)&=
    \left(\left\lfloor\frac{j+k}{2}\right\rfloor,
       \left\lceil\frac{j+k}{2}\right\rceil\right),\\
    Q_{j_*k_*}(\bm s_a,\bm s_b)
    &=-\frac12\left[
    Q_{jk}(\bm s_a,\bm s_b)+Q_{jk}(\bm s_b,\bm s_a)
    \right].
    \end{aligned}
    \label{eq:supp_weighted_center_completion}
\end{equation}
This polynomial is symmetric and has degree $r$ in each spinor,
so it is allowed at either a same-orbital or a neighboring-orbital center
pair within the same LL cutoff. It gives
\begin{equation}
    Q_{jk}(\bm s,\bm s)+Q_{j_*k_*}(\bm s,\bm s)=0.
    \label{eq:supp_weighted_squeezed_cancellation}
\end{equation}
The center pair is strictly inward of $(j,k)$, which remains the root.
Consequently, these more distant pairs impose no additional direct
bracket factor on the root coefficient.

To express these conditions as operator constraints, let $c_{j,\alpha}$
annihilate a particle in internal state $\alpha$ at orbital $j$,
in an orthonormal single-particle basis. Define
\begin{equation}
    Z_{jk,\lambda}
    =\sum_{\alpha,\beta=0}^{r}
    z_{\alpha\beta}^{jk,\lambda}c_{j,\alpha}c_{k,\beta},
    \qquad\lambda=0,\ldots,2r.
    \label{eq:supp_weighted_Z_definition}
\end{equation}
The coefficients $z_{\alpha\beta}^{jk,\lambda}$ are fixed by requiring
\begin{equation}
    Z_{jk,\lambda}|\psi_{jk}\rangle
    =\left([u^{2r-\lambda}v^\lambda]\,
    Q_{jk}(\bm s,\bm s)\right)|0\rangle,
    \label{eq:supp_weighted_Z_action}
\end{equation}
for every two-particle state $|\psi_{jk}\rangle$ with occupation $(j,k)$. Here the square brackets denote coefficient extraction, and $Q_{jk}$ is the polynomial of that state in the convention above. This definition includes the change to the polynomial basis and its normalization factors.

For $k=j,j+1$, the two-particle analysis gives
$Z_{jk,\lambda}|\psi_{jk}\rangle=0$ whenever $(j,k)$ is a root of a
contact zero mode. The same operator annihilates every other two-particle occupation because it removes particles only from $j$ and $k$.
Thus $Z_{jk,\lambda}$ annihilates the root components of every
two-particle zero mode.

The root theorem extends this result to many particles
\cite{CruiseSeidel2023DNA}. To apply the theorem, we annihilate the other $N-2$ particles with $\mathcal C=\prod_{\ell=1}^{N-2}c_{j_\ell,\alpha_\ell}$. The pure annihilation operators
$\mathcal A_\nu$ defining the contact zero-mode conditions satisfy
\begin{equation}
    \mathcal A_\nu\mathcal C|\Psi_N\rangle
    =\mathcal C\mathcal A_\nu|\Psi_N\rangle=0.
    \label{eq:supp_weighted_spectator_contraction}
\end{equation}
Moreover, a nonzero pair component inherited from an original root
remains a root after contraction: a more outward contribution after
contraction would imply a more outward occupation in the original state.
For an $N$-particle contact zero mode $|\Psi_N\rangle$ and one of its
root occupation patterns $\boldsymbol n=(n_j)$, let $\Pi_{\boldsymbol n}$
denote the projector onto the subspace with $n_j$ particles in each
guiding-center orbital $j$, retaining all internal LL configurations.
The theorem therefore gives
\begin{equation}
    Z_{jk,\lambda}\Pi_{\boldsymbol n}|\Psi_N\rangle=0,
    \qquad k=j,j+1,\quad\lambda=0,\ldots,2r.
    \label{eq:supp_weighted_root_theorem}
\end{equation}
In the polynomial representation, these constraints imply
\begin{equation}
    Q_{\boldsymbol n}
    =\left(
    \prod_{\substack{a<b\\j_a=j_b}}[ab]^2
    \prod_{\substack{a,b\\j_b=j_a+1}}[ab]
    \right)R_{\boldsymbol n},
    \label{eq:supp_weighted_manybody_factors}
\end{equation}
where $j_a$ is the orbital occupied by particle $a$, and
$R_{\boldsymbol n}$ is symmetric among particles in each orbital.

For each particle at an occupied orbital $j$, the same-orbital factors use
$2(n_j-1)$ degrees, and the neighboring-orbital factors use
$n_{j-1}+n_{j+1}$. The remaining degree is therefore
\begin{equation}
 d_j=n_{\rm LL}-1-2(n_j-1)-n_{j-1}-n_{j+1}.
 \label{eq:supp_weighted_residual_degree}
\end{equation}
For $d_j\geq0$, the remaining single-particle polynomials span a
$(d_j+1)$-dimensional space. Bosonic symmetry within orbital $j$ gives
\begin{equation}
 f_j(\boldsymbol n)
 =\dim\operatorname{Sym}^{n_j}\!\left(\mathbb C^{d_j+1}\right)
 =\binom{d_j+n_j}{n_j},\qquad n_j>0,\quad d_j\geq0.
 \label{eq:supp_weighted_local_dimension}
\end{equation}
The residual coefficient spaces at different orbitals form a tensor
product. The weight of an occupation pattern is thus
$\prod_j f_j(\boldsymbol n)$. With periodic orbital indices, the EPP counting rule predicts the torus zero-mode count
\begin{equation}
\begin{aligned}
 D_{\rm EPP}
 &=\sum_{\sum_j n_j=N_A}\prod_{j=0}^{L-1}f_j(\boldsymbol n),\\
 f_j(\boldsymbol n)
 &=\begin{cases}
 1,&n_j=0,\\[2pt]
 \displaystyle\binom{n_{\rm LL}+1-n_{j-1}-n_j-n_{j+1}}{n_j},
 &n_j>0,\quad 2n_j+n_{j-1}+n_{j+1}\leq n_{\rm LL}+1,\\[4pt]
 0,&\text{otherwise}.
 \end{cases}
\end{aligned}
\label{eq:supp_weighted_epp}
\end{equation}
This formula counts the independent root states allowed by the local constraints. For the continuum bosonic $21$ and $22$ models on the disk, every root state allowed by the respective EPP can be completed to a many-body zero mode~\cite{Bandyopadhyay2020Jain,Bandyopadhyay2018EPP,CruiseSeidel2026}. We conjecture that the same holds for the $33$ model. To predict the zero-mode count on the torus, we further assume that the same local weights apply with periodic orbital indices. We test this prediction against the finite-size lattice spectra below.

\paragraph{Momentum resolution.}
At flux $\phi=p/q$, each LL contains $L=N_\phi=p\mathcal N_x\mathcal N_y$ guiding-center orbitals. We choose the basis $|j\rangle$ so that the single-particle magnetic translations $t_x$ and $t_y$ act as follows~\cite{BernevigRegnault2012Momentum}:
\begin{equation}
    t_x|j\rangle=|j+p\mathcal N_y\rangle,\qquad
    t_y|j\rangle=e^{-2\pi ij/\mathcal N_y}|j\rangle,\qquad
    K_y^A=\sum_j jn_j\pmod{\mathcal N_y}.
    \label{eq:MP_guiding_center_translations}
\end{equation}
Orbital indices are taken modulo $L$; internal LL factors translate
with their roots. Character decomposition using
Eq.~\eqref{eq:MP_translation_sectors} gives $(114,108,108)$ for $H_{21}$
and $(138,138)$ for $H_{22}$ at $p>1$, matching the PES in every sector.
\subsection{Finite-size spectra and counting comparisons}
\label{app:numerical_results}

Table~\ref{tab:numerical_counts} compares the finite-size PES ranks
with the independent counting benchmarks specified in its last column.

\begin{center}
\begin{minipage}{\linewidth}
\captionof{table}{
PES support and counting comparisons for
Figs.~\ref{fig:parton_supp}--\ref{fig:rr_supp}.
Sector counts are ordered by increasing $K_A$ and agree with the
independent comparison specified in the last column.}
\label{tab:numerical_counts}
\label{tab:weighted_counts}
\label{tab:h33_support}
\label{tab:rr_counts}
\centering\small
\setlength{\tabcolsep}{4pt}
\renewcommand{\arraystretch}{1.13}
\begin{tabular}{lccrll}
\hline
Model & $\phi$ & $N_A$ & Total & Sector counts & Comparison\\
\hline
$H_{21}$ & $1/4$ & 3 & 330 &
$(42,36,36,36,36,36,36,36,36)$ & EPP\\
$H_{22}$ & $1/6$ & 3 & 276 &
$(48,45,45,48,45,45)$ & EPP\\
$H_{33}$ & $1/12$ & 4 & 400 &
$(108,96,100,96)$ & Independent model-state PES\\
$H_{33}$ & $1/12$ & 3 & 388 &
$(97,97,97,97)$ & EPP\\
\hline
$H_{21}$ & $3/10$ & 3 & 330 &
$(114,108,108)$ & EPP\\
$H_{22}$ & $3/16$ & 3 & 276 &
$(138,138)$ & EPP\\
\hline
Spin-1 MR & $1/2$ & 4 & 170 &
$(26,20,24,20,20,20,20,20)$ & $(2,2)$ GPP\\
Spin-$3/2$ $Z_3$ RR & $1/2$ & 4 & 102 &
$(18,18,18,16,16,16)$ & $(3,2)$ GPP\\
Dressed MR & $3/4$ & 4 & 4758 &
$(618,588,612,588,588,588,588,588)$ & GPP [Eq.~\eqref{eq:supp_dressed_gpp}]\\
\hline
\end{tabular}
\end{minipage}
\end{center}

\begin{figure}[!p]
    \centering
    \includegraphics[width=\linewidth]{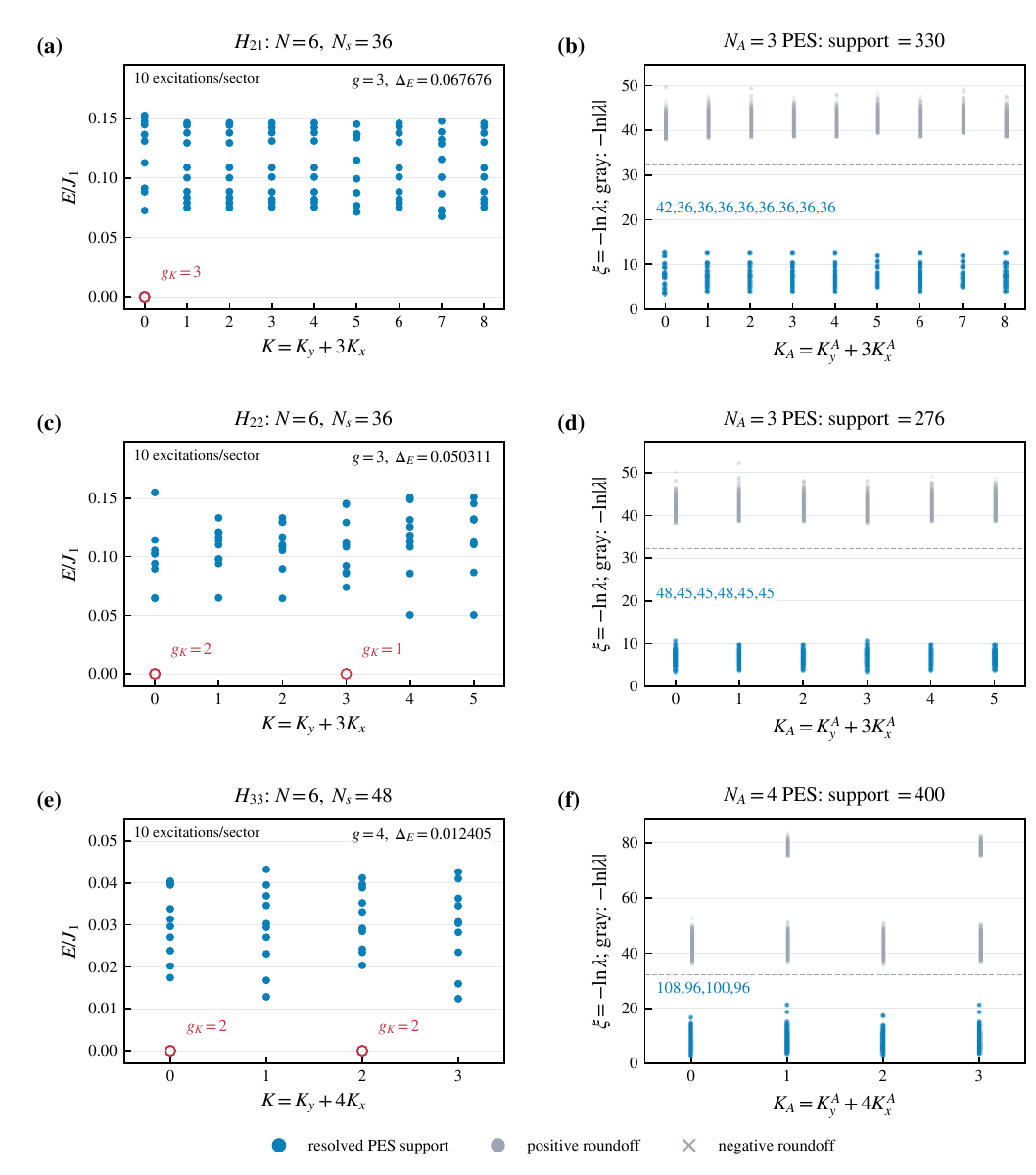}
    \caption{\textbf{Energy spectra and PES for $H_{mn}$ at $p=1$.} The rows show $H_{21}$, $H_{22}$, and $H_{33}$ (Table~\ref{tab:numerical_parameters}). In the left panels, the red labels $g_K$ give the number of zero modes in each momentum sector. Energies are given in units of $J_1=(|\langle\bm a_1|h|\bm0\rangle|+|\langle\bm a_2|h|\bm0\rangle|)/2$, where $h$ is the finite-torus single-particle Hamiltonian. The right panels show the momentum-resolved PES. The level counts agree with the EPP prediction for $H_{21}$ and $H_{22}$ and with independent counts from the parton wavefunctions for $H_{33}$ (Table~\ref{tab:numerical_counts}). For $H_{33}$, all nonzero PES eigenvalues are shown, while additional zeros of the full reduced density matrix are omitted.}
    \label{fig:parton_supp}
\end{figure}

\begin{figure}[!p]
    \centering
    \includegraphics[width=\linewidth]{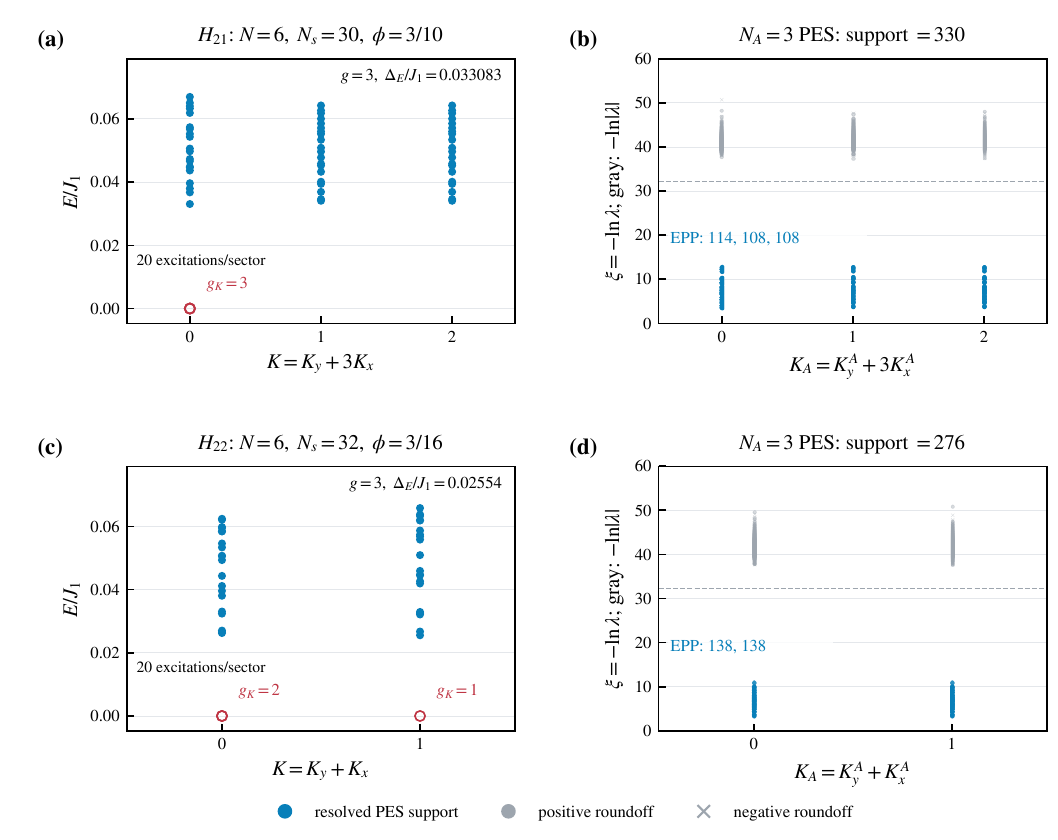}
    \caption{\textbf{Energy spectra and PES for $H_{mn}$ at $p>1$.} The rows show $H_{21}$ at $\phi=3/10$ and $H_{22}$ at $\phi=3/16$. In the left panels, the red labels give the number of zero modes in each momentum sector. Energies are given in units of $J_1$, as defined in Fig.~\ref{fig:parton_supp}. The right panels show the $N_A=3$ PES. The number of nonzero levels in each momentum sector agrees with the EPP prediction (Table~\ref{tab:numerical_counts}).}
    \label{fig:parton_rational}
\end{figure}

\begin{figure}[!p]
    \centering
    \includegraphics[width=\linewidth]{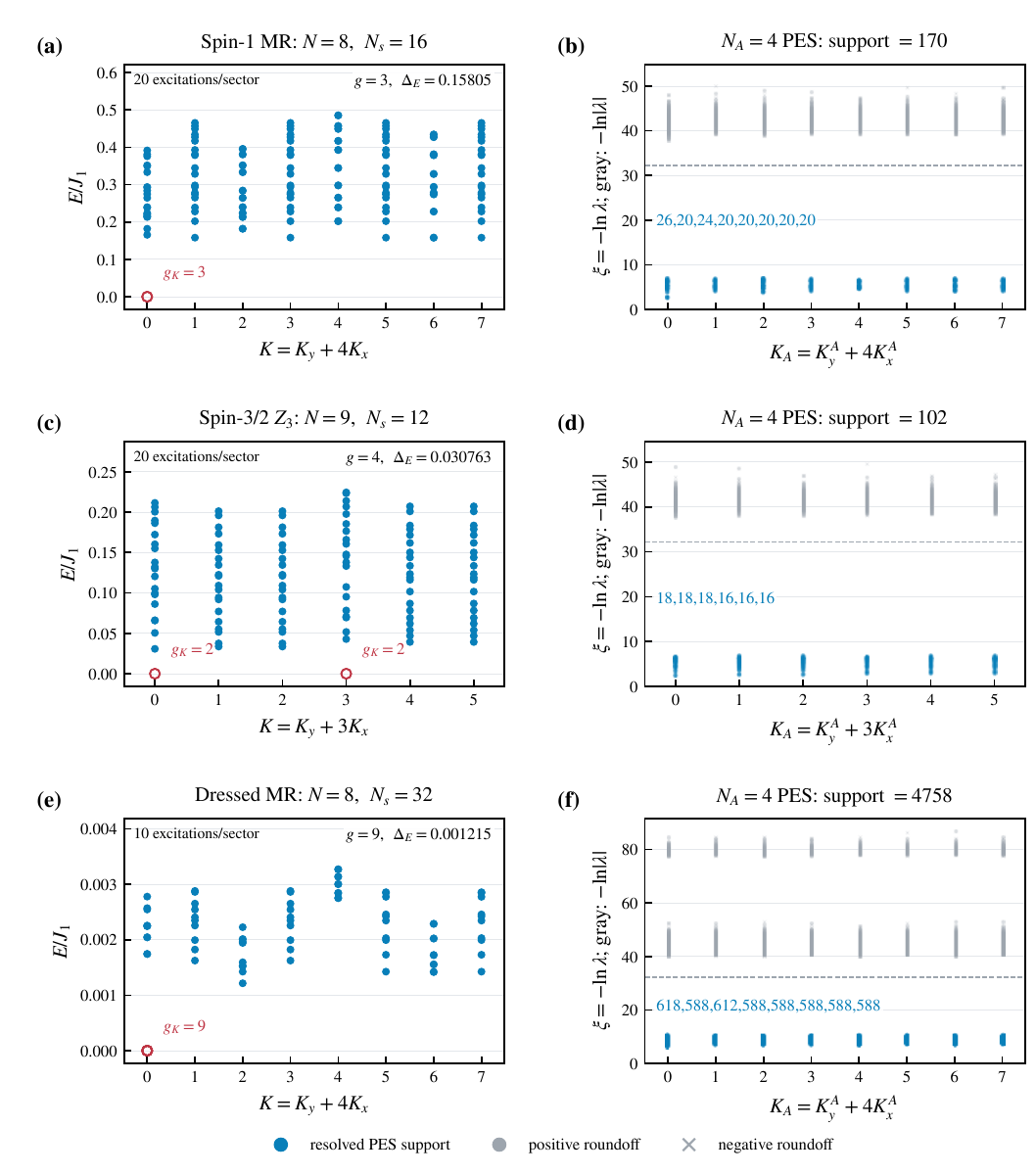}
\caption{\textbf{Energy spectra and PES of the Read--Rezayi and dressed Moore--Read models.} In the left panels, the red labels $g_K$ give the number of zero modes in each momentum sector. Energies are given in units of $J_1$, the nearest-neighbor hopping magnitude for each model. Panel (e) shows the dressed Moore--Read spectrum in the limit of infinite three-site repulsion. The right panels show the $N_A=4$ PES. The number of nonzero levels in each momentum sector agrees with the GPP count in Table~\ref{tab:numerical_counts}.}
    \label{fig:rr_supp}
\end{figure}

For the numerical results in Fig.~\ref{fig:highchern54-spectrum-pes},
we use the three-sublattice construction of
Sec.~\ref{sec:HC_single_band} with
$h_{\bm k}^{(C)}=J_C Q_{\bm k}^{\rm LL}$
[Eq.~\eqref{eq:HC_parent}], $C=-2$, and $J_C>0$.
The angle between the primitive vectors $\bm a_1$ and $\bm a_2$ is $120^\circ$. The three sublattice positions are
$\bm\tau_A=(2\bm a_1+\bm a_2)/3$,
$\bm\tau_B=(\bm a_1-\bm a_2)/3$, and $\bm\tau_O=\bm0$,
with $(\mathcal B_A,\mathcal B_B,\mathcal B_O)
=(e^{-i\pi/3},-1,e^{i\pi/3})$. We set $\bm\rho_0=\bm0$.
We use a $6\times3$ torus of primitive cells with $N_s=54$ sites and zero boundary twists. The sector with six up spins ($N=6$) has magnetization $m_0=-7/9$ and corresponds to bosonic filling $\nu_b=1/3$.

\begin{figure}[!htp]
    \centering
    \includegraphics[width=.49\linewidth]{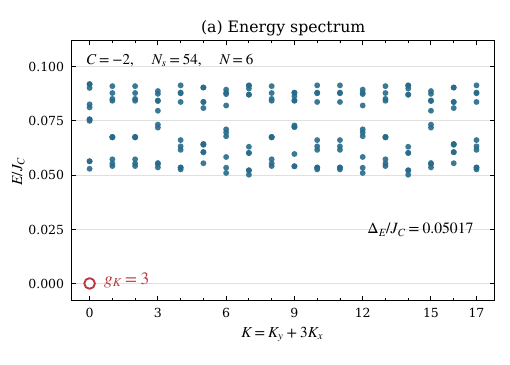}%
    \hfill
    \includegraphics[width=.49\linewidth]{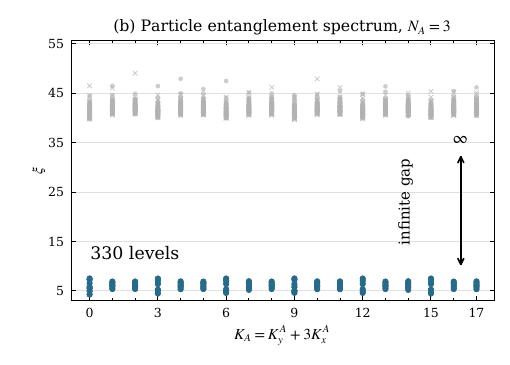}
    \caption{\textbf{Spin-$1/2$ model for a Halperin $(221)$-type chiral spin liquid.}
    (a) Energy spectrum of the full spin Hamiltonian in the sector with six up spins. All three zero modes lie at $(K_x,K_y)=(0,0)$. In each momentum sector, we show the ten lowest positive-energy states and any additional states degenerate with the tenth.
    The finite-size separation is $\Delta_E/J_C\simeq0.05017$.
    (b) $N_A=3$ PES of the equal-weight mixture of the three zero modes,
    evaluated in the flatband orbital basis. Its 330 nonzero eigenvalues
    comprise 21 in each of sectors $(0,0)$ and $(3,0)$ and 18 in each
    remaining sector. Gray circles and crosses mark small positive and negative eigenvalues, respectively, that are treated as numerical zeros.
    Both panels use Eq.~\eqref{eq:MP_translation_sectors} with
    $(\mathcal N_x,\mathcal N_y)=(6,3)$.}
    \label{fig:highchern54-spectrum-pes}
\end{figure}

\paragraph{Quasihole sector of $H_{33}$.}
At $N_\phi=4$, the $\nu_b=3/2$ parton state has six particles. We keep the flux fixed and count the zero modes of $H_{33}$ in the four-particle quasihole sector.
We use the same $6\times8$, $\phi=1/12$ torus and sampling convention
as in Fig.~\ref{fig:parton_supp}, with zero twists.
We diagonalize the full spin Hamiltonian in the sector with four up spins on 48 sites. In order of increasing momentum sector, we find $(108,96,105,96)$ zero modes, totaling 405. These counts agree sector by sector with the EPP prediction and with an independent count of four-boson states built from the 20 zero-energy single-particle orbitals subject to the onsite hard-core constraint.

The $N_A=4$ PES of the six-particle ground-state manifold has sector counts $(108,96,100,96)$, totaling 400. The sampled $\chi_3^2$ wavefunctions give the same PES counts in every momentum sector. The four-particle PES therefore spans 400 of the 405 zero-mode dimensions, with all five missing dimensions in sector $K_A=2$.
The $N_A=3$ row of Table~\ref{tab:numerical_counts} refers to the three-particle cut of the same six-particle ground manifold.

\begin{figure}[!htp]
    \centering
    \includegraphics[width=4in]{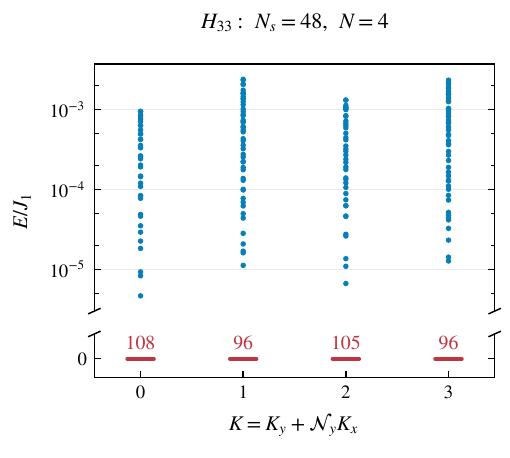}
    \caption{\textbf{Energy spectrum of $H_{33}$.} We show the lowest 144 levels in each momentum sector. Red labels give the number of zero modes in each sector, totaling 405. Positive energies are shown on a logarithmic scale. The smallest resolved positive energy is
    $E/J_1\simeq4.71\times10^{-6}$.
    The energy scale $J_1$ is defined as in Fig.~\ref{fig:parton_supp}.
    Momenta follow Eq.~\eqref{eq:MP_translation_sectors} with
    $(\mathcal N_x,\mathcal N_y)=(1,4)$.}
    \label{fig:h33_n4_hardcore}
\end{figure}

\endgroup
\end{document}